\documentclass[12pt,twoside]{book}

\usepackage{rscsoftmat}
\usepackage{xcolor}
\newcommand\blue[1]{ {\color{blue}#1} }
\newcommand\red[1]{ {\color{red}#1} }
\definecolor{pymagenta}{HTML}{C20078}
\newcommand\magenta[1]{ {\color{pymagenta}#1} }

\usepackage[usetitle]{rsc}

\newcommand{\chapterauthor}{Jérémy O'Byrne$^a$, Alexandre Solon$^b$, Julien Tailleur$^{a,\star}$, Yongfeng Zhao$^{c}$} %authors
\newcommand{\authoraffiliation}{$^a$Universit\'e de Paris, Laboratoire Matière et Systèmes Complexes (MSC), UMR 7057 CNRS, F-75205 Paris, France, $^b$Sorbonne Université, CNRS, Laboratoire Physique Théorique de la Matière Condensée, 75005 Paris, France, $^c$ Center for Soft Condensed Matter Physics and Interdisciplinary Research \& School of Physical Science and Technology, Soochow University, 215006 Suzhou, China} %affiliation and adress
\newcommand{\email}{julien.tailleur@univ-paris-diderot.fr} % email address of corresponding author

\newcommand{\abstract}{
Active particles may undergo phase separation when interactions oppose self-propulsion,
in the absence of any cohesive forces. The corresponding Motility-Induced Phase Separation (MIPS) is arguably the simplest non-trivial collective feature that distinguishes active from passive particles. It is observed in a large variety of systems which we review in this chapter. We describe in depth the case of motile particles interacting via quorum-sensing interactions, whose theoretical framework is by now well-established. We close the chapter by discussing the features observed in systems undergoing MIPS that still challenge our understanding.}

\newcommand\bfr{{\bf r}}
\newcommand\bfu{{\bf u}}
\newcommand\bfJ{{\bf J}}
\newcommand\bfm{{\bf m}}
\newcommand\bfi{{\bf i}}
\newcommand\C{{\mathcal{C}}}
\newcommand\bfF{{\bf F}}
\newcommand\grad{\boldsymbol\nabla}
\usepackage{hyperref}
\usepackage{tikz}
\usepackage{bm}
\usepackage{stmaryrd}

\begin{document}

\chapter{An Introduction to Motility-Induced Phase Separation}

\MiniToC

\section{Introduction}
\label{sec:intro}

The liquid-gas phase separation of simple fluids is arguably
the simplest phase transition encountered in equilibrium systems. Its
microscopic origin can be understood from a simple inspection of the
Boltzmann distribution. The latter implies that the most likely
macroscopic state in which a system is found results from a
competition between its entropy and its energy. In the presence of
attractive interactions between the molecules, energy favours cohesion
whereas entropy favors disorder. The relative weights of entropy and
energy in the steady state are controlled by temperature. When the
latter is lowered, energy wins over entropy, promoting the emergence
of a dense fluid phase over a gaseous state. When the total
number of particles is conserved, this phase transition is replaced by
phase coexistence.

Consider a microscopic model of attractively interacting particles in equilibrium. While the origin of its phase separation can be qualitatively accounted for by inspection of the Boltzmann weight, predicting any macroscopic property of the system starting from the microscopic level is a
fantastic challenge. Instead, one rather coarse-grains the system to
describe it at a mesoscopic level, using the so-called model B
dynamics~\cite{hohenberg_theory_1977}
\begin{equation}\label{eq:modelB}
  \frac{\partial }{\partial t}\rho(\bfr,t)=\nabla\cdot\left[\Gamma_0 \nabla\frac{\delta \mathcal F[\rho]}{\delta \rho(\bfr)}+\boldsymbol\eta(\bfr,t)\right]
\end{equation}
where $\mathcal F[\rho]$ is the free energy of the coarse-grained
configuration $\rho(\bfr,t)$, $\Gamma_0$ is a mobility constant, and
$\boldsymbol\eta(\bfr,t)$ is a Gaussian white noise with variance
$\langle\eta_\alpha(\bfr',t')\eta_\beta(\bfr,t)\rangle=\Gamma_0\delta_{\alpha\beta}\delta(t-t')\delta^d(\bfr-\bfr')$. Equation~\eqref{eq:modelB}
describes the evolution at late times and large scales of the order
parameter of the system, here the density field of the fluid, and
allows predicting the universal properties of equilibrium liquid-gas
phase separation. Mean-field theories built out of
Eq.~\eqref{eq:modelB} also accurately predict its phase
diagram~\cite{chaikin1995principles}, outside a Ginzburg interval
around the critical point.

\begin{figure}
  \includegraphics[width=\textwidth]{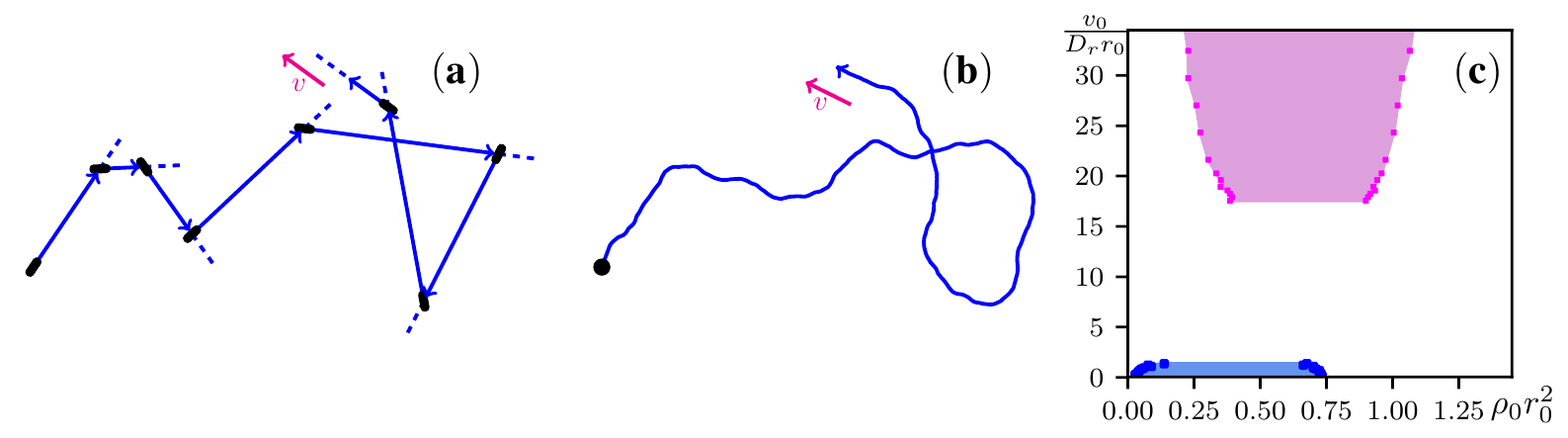}
  \if{
    \begin{tikzpicture}
      \path (0,0) node {\raisebox{.5cm}{\includegraphics[width=.65\textwidth]{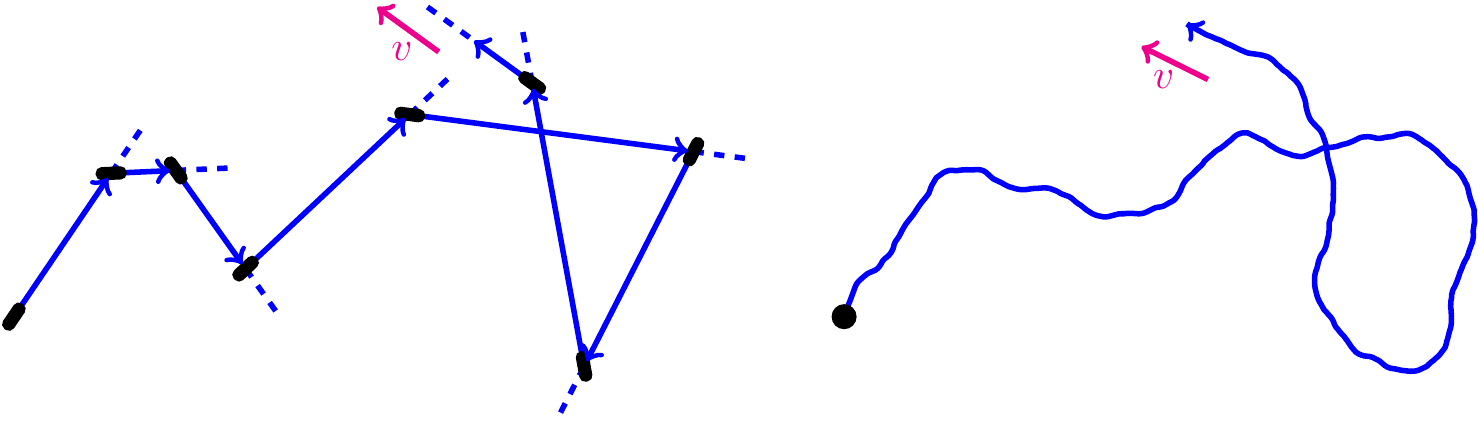}}}; 
      \path (8,0) node {\includegraphics[]{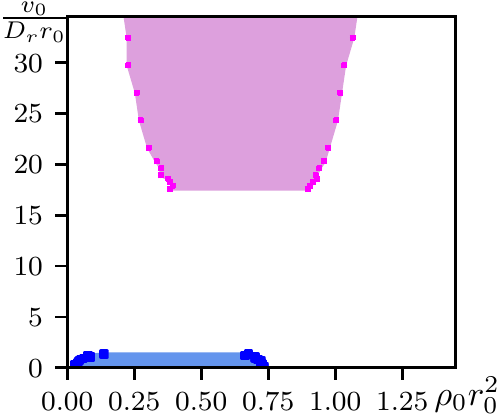}};
      \draw (-.75,1.5) node {$\bf (a)$};
      \draw (4.5,1.5) node {$\bf (b)$};
      \draw (9.75,1.5) node {$\bf (c)$};
  \end{tikzpicture}}
  \fi
  \caption{{\bf (a)} Schematic trajectory of a Run-and-Tumble Particle (RTP). Runs in straight lines at fixed speed $v$ are punctuated by tumbles, occurring with rate $\alpha$, during which the particle randomizes its orientation. {\bf (b)} Schematic trajectory of an Active Brownian Particle (ABP). The orientation diffuses continuously, with a rotational diffusivity $D_r$. {\bf (c)} Phase diagram of self-propelled ABPs interacting via
    a Lennard-Jones pair potential (See~\protect\cite{o2021time} for details of the model). The equilibrium phase-separated region (blue) survives for small enough self-propulsion speeds. Motility-induced phase separation (plum) is observed whenever the run-length $v_0/D_r$ is significantly larger than $15$ particle diameters $r_0$. Symbols correspond to coexisting densities. }\label{fig:MIPSLJ}
\end{figure}
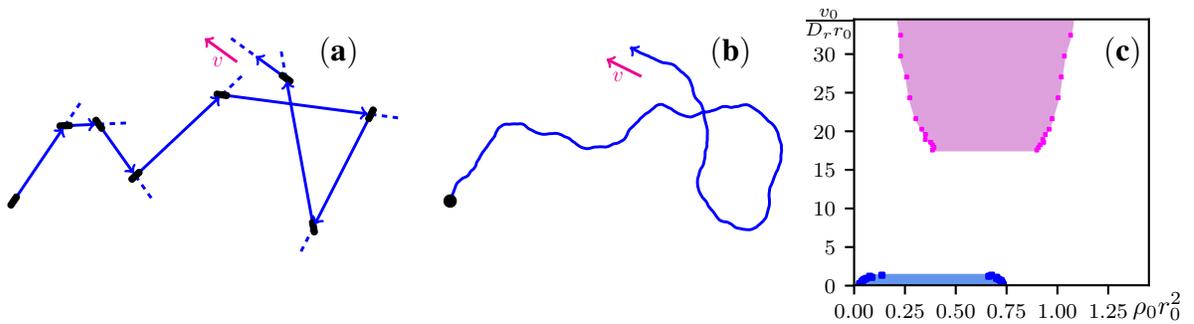

In active matter, none of the tools described above are
available to predict the phase behaviors of simple active fluids. Let
us consider the persistent dynamics illustrated in
Fig~\ref{fig:MIPSLJ} which model the two-dimensional motion of
bacteria and self-propelled colloids. A natural question is whether
replacing the Brownian dynamics of passive colloids by such
self-propelled dynamics completely alters the understanding of
phase separation inherited from equilibrium physics. In the
weak propulsion limit, or when attractive forces are strong enough to
overcome self-propulsion ones, one expects the qualitative features of the passive
case to survive. At larger self-propulsion speeds, active forces are likely to overcome attractive ones, leading to homogeneous phases. Interestingly, Fig.\ref{fig:MIPSLJ} shows a
richer picture. While phase-separation survives in the
weak-self-propulsion, near-equilibrium regime, it is mostly observed far from it, when the
persistence length of the particles is much larger than their
individual size. Even more surprisingly, this phase separation would also be observed in the
complete absence of attractive interactions. The underlying mechanism,
known as motility-induced phase separation~\cite{cates2015motility}
(MIPS), is the topic of this chapter.

In Section~\ref{sec:phenomenology}, we start by reviewing the variety
of systems in which MIPS has been reported. In
Section~\ref{sec:handwaving}, we show that it emerges from an
instability due to a simple feedback mechanism. First, active
particles generically accumulate where they go slower. Consequently,
when interactions make active particles slow down at high density,
they tend to accumulate where they are already denser than the
average, hence making homogeneous phases unstable to density
fluctuations. In Section~\ref{sec:vofr}, we demonstrate this tendency
to accumulate in slower regions and characterize the large-scale
long-times dynamics of active particles endowed with a non-uniform
speed $v({\bf r})$. The latter is then used in Section~\ref{sec:QSAPs}
to build a large-scale description of active particles interacting via
quorum-sensing interactions. This theory can then be used to predict
MIPS for this systems, and to characterize the corresponding phase
diagram, which is the topic of Section~\ref{sec:generalized
  thermo}. In Section~\ref{sec:PFAPs}, we then discuss the case of
active particles interacting via pairwise forces, highlighting both
similarities and differences with the case of quorum sensing. In
Section~\ref{sec:lattice}, we finally discuss MIPS in lattice
models. We close this chapter in Section~\ref{sec:conclu}
by discussing open problems and current research topics on MIPS.

\section{Phenomenology}
\label{sec:phenomenology}
Let us first describe the types of systems that undergo MIPS. As we
will show in the next sections, MIPS is triggered by a positive
feedback loop: If interactions act to slow down the active particles,
a small aggregate will accumulate more and more particles to
ultimately lead to phase separation. Two types of interacting active particles have
been shown, both in models and experiments, to experience such a slowdown leading to MIPS:
\begin{itemize}
\item \emph{Quorum-Sensing Active Particles (QSAPs)}, which interact by
  adapting their velocity to the local density. Such quorum-sensing
  interactions are displayed, for example, by bacteria which may change their motility in response to the local
  concentration of a chemical that they release in the environment.  This
  type of interaction, for which MIPS was first
  predicted~\cite{tailleur_statistical_2008}, is best amenable to
  analytic treatment. In particular, we will show in
  Sec.~\ref{sec:QSAPs} that mesoscopic equations can be derived for such models using an adequate
  coarse graining procedure.
\item \emph{Pairwise-Forces Active Particles (PFAPs)} which interact
  via short-range pairwise repulsive interactions such as excluded
  volume interactions. Indeed, two active particles colliding
  head-to-head will stall each other until one of them either rotate or slide along
  the other. The speeds of the particles are thus reduced during the
  collision. When this effect is strong enough, it may trigger MIPS. This
  type of particles, discussed in more detail in Sec.~\ref{sec:PFAPs},
  is harder to study analytically than QSAPs and offers a richer
  phenomenology, several  aspects of which are still debated.
\end{itemize}

\begin{figure}
  \includegraphics[width=0.48\textwidth]{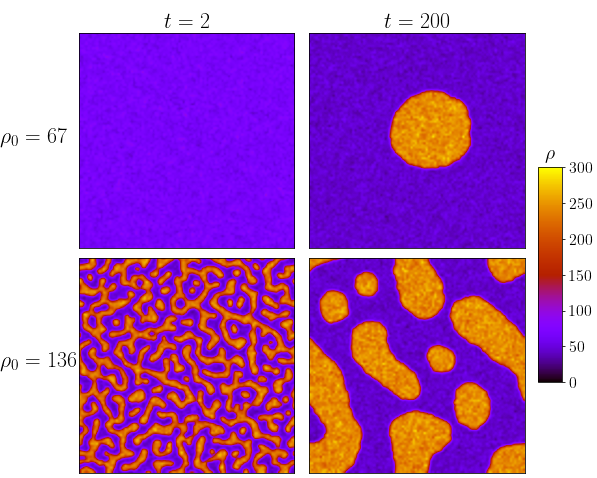}
  \includegraphics[width=0.48\textwidth]{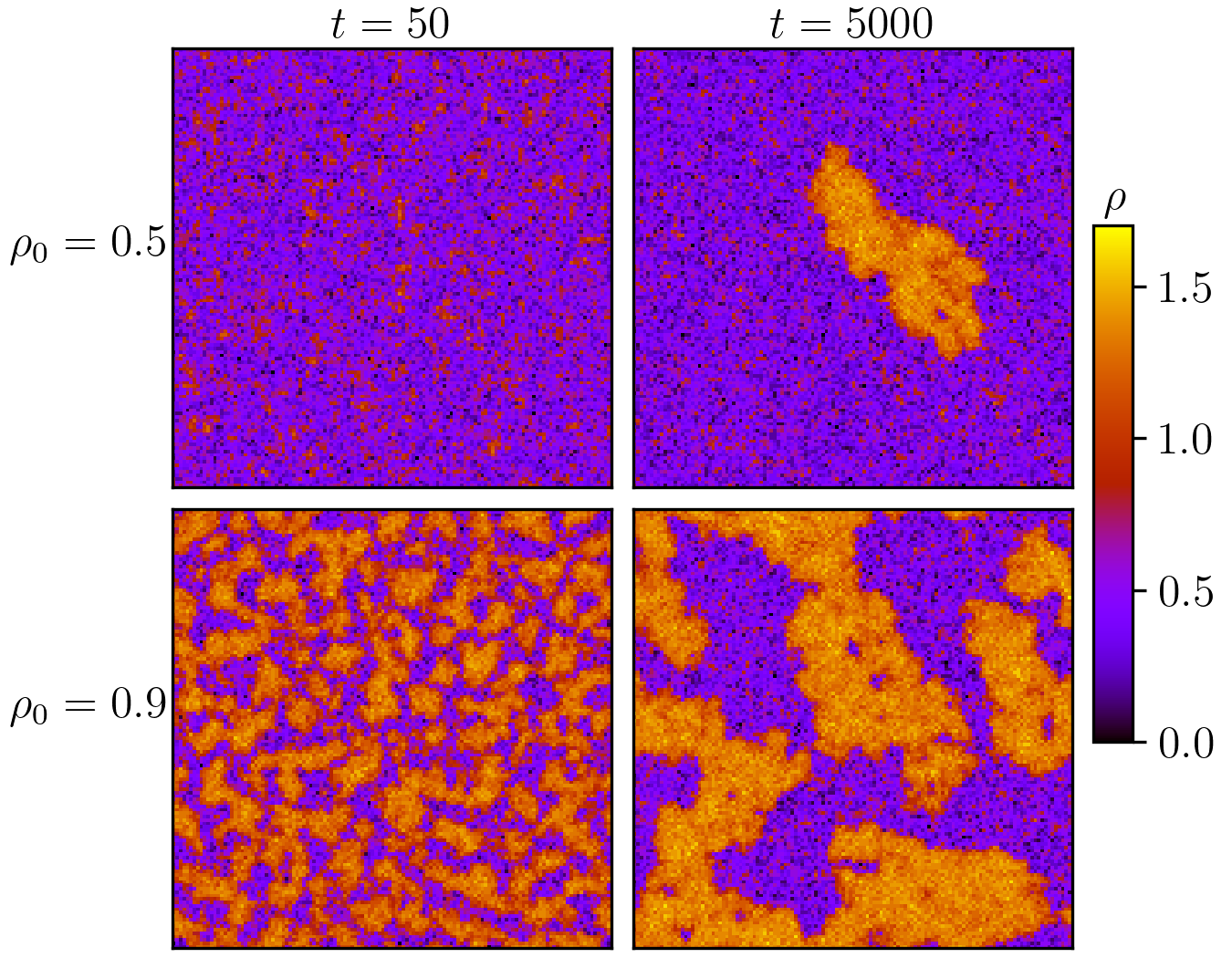}
  \caption{Nucleation (top rows) and spinodal decomposition (bottom rows). Left: QSAPs, see Fig~.1 of Ref.~\cite{solon2018PRE} for the details of the model. Right: PFAPs with the interaction potential $V(r)=k(\sigma-r)^2\Theta(\sigma-r)$, $k=50$, $\sigma=1$, $v=5$, $D_r=0.2$.}\label{fig:snapshots}
\end{figure}

Both QSAPs and PFAPs come in different flavors corresponding to
different dynamics of the self-propulsion velocity. Two common self-propulsion dynamics are shown
in Fig.~\ref{fig:MIPSLJ}: Active Brownian particles (ABPs),
which reorient continuously because of rotational noise, and
run-and-Tumble Particles (RTPs), which ``run'' in straight lines
between two ``tumbles'' during which they randomize their
orientations. Active Ornstein-Uhlenbeck Particles (AOUPs)~\cite{szamel2014self,martin_statistical_2021} are yet another
kind of particles which have the interesting property of having
Gaussian fluctuations of their self-propulsion velocity. Although the
trajectories of these three particles seem markedly different, they all correspond to persistent random walks. All undergo MIPS and
current knowledge suggests that these different active dynamics do not lead to important
qualitative
differences~\cite{solon2015pressure,martin_statistical_2021}.

In all these models, MIPS is observed with a phase diagram similar to the large velocity region of
Fig.~\ref{fig:MIPSLJ}c. Between the two binodal lines, one observes a
phase-separated steady state where a high density region comprising slowly
moving particles coexists with a gaseous phase in which particles move more rapidly. Moreover, this coexistence state bears the hallmarks of liquid-gas phase separation
that are spinodal decomposition, nucleation and metastability. Inside
but close to the binodals, a homogeneous system is metastable until a
large enough fluctuation nucleates a droplet and brings the system to
a phase coexistence. On the contrary, between the spinodal lines, any
homogeneous system is unstable to arbitrarily small fluctuations.  One then observes spidonal
decomposition: Starting from a homogeneous profile, many liquid or
gas clusters are formed (depending on the majority phase), which then
coarsen in time. Both phenomena are illustrated in
Fig.~\ref{fig:snapshots}.

\begin{figure}
  \includegraphics[width=\textwidth]{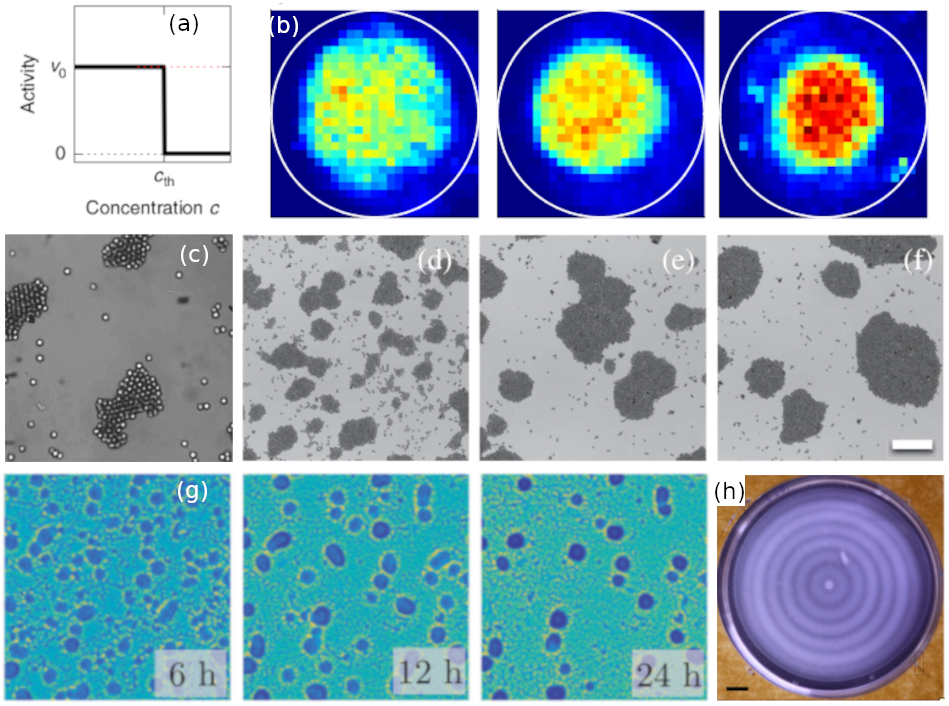}
  \caption{Experimental realizations of MIPS. (a,b) Self-propelled
    colloids whose motility is tuned by a computer-assisted loop to
    obtain the desired quorum-sensing interactions (a). One observes
    phase separation (b) with warmer colors denoting higher
    densities. The three snapshots correspond to three values of
    threshold concentration $c_{\rm th}$ increasing from left to
    right. (c) Self-propelled colloids interacting via pairwise repulsive
    forces. (d-f)
    Self-propelled colloids showing coarsening (d,e) up to a certain
    length scale (e,f) where MIPS is arrested because of alignment
    inside the clusters. (g) Formation of
    fruiting bodies for bacteria \emph{Myxococcus
      xanthus}. (h) Arrested MIPS in
    genetically engineered \emph{E. coli} bacteria where the
    population dynamics in the colony compete with MIPS to select a
    characteristic pattern size. Panel (a,b) are reproduced from Ref.~\cite{bauerle_self-organization_2018} under the \href{https://creativecommons.org/licenses/by/4.0/}{Creative Commons licence}. Panel (c) is reproduced with permission from Ref.~\cite{buttinoni_dynamical_2013} \href{https://doi.org/10.1103/PhysRevLett.110.238301}{DOI}. [I. Buttinoni, J. Bialké, F. Kümmel, H. Löwen, C. Bechinger and T. Speck, Physical review letters, 110, 238301, 2013]. Copyright (2013) by the
American Physical Society. Panel (d,f) are reproduced with permission from Ref.~\cite{van_der_linden_interrupted_2019} \href{https://doi.org/10.1103/PhysRevLett.123.098001}{DOI}. [M. N. Van Der Linden, L. C. Alexander, D. G. Aarts and O. Dauchot, Physical review letters,123, 098001, 2019]. Copyright (2019) by the
American Physical Society. Panel (g) is reproduced with permission from Ref.~\cite{liu_self-driven_2019} \href{https://doi.org/10.1103/PhysRevLett.122.248102}{DOI}. [G. Liu, A. Patch, F. Bahar, D. Yllanes, R. D. Welch, M. C. Marchetti, S. Thutupalli and J. W. Shaevitz, Physical review letters,122, 248102, 2019]. Copyright (2019) by the
American Physical Society. Panel (h) is courtesy of Prof Jian-Dong
    Huang.}\label{fig:manips}
\end{figure}

Experimentally, MIPS has been investigated in different systems,
either living (bacteria) or inert (self-propelled colloids). The
most clear-cut realization of MIPS is arguably that
of Ref.~\cite{bauerle_self-organization_2018} which uses micrometric silica beads half-coated
with carbon in a water-lutidine mixture. The activity of each particle
is tuned in real time by using a laser focusing on the particles to adapt their
velocities on the local density (see Fig.~\ref{fig:manips}a), as for
QSAPs \textit{in silico}. This triggers MIPS, as shown in Fig.~\ref{fig:manips}b. The
same self-propelled colloids, when endowed with a uniform self-propulsion speed and interacting
via pairwise repulsive forces, are found to form large clusters (see
Fig.~\ref{fig:manips}c) that have also been described as an experimental realization of
MIPS~\cite{buttinoni_dynamical_2013}.

In many other experimental systems, clustering or phase separation has been reported but a clear connection to MIPS proved elusive. First, additional ingredients may come in and arrest a \textit{bona fide} MIPS. This is for instance the case for the
titanium-coated Janus colloids that are self-propelled by  an AC
external electric field~\cite{van_der_linden_interrupted_2019}. Phase separation is found to be arrested
(see Fig.~\ref{fig:manips}d,e,f), possibly because of the aligning interactions experienced by the colloids in the dense phase. Hydrodynamic~\cite{matas-navarro_hydrodynamic_2014,zottl_hydrodynamics_2014} and phoretic~\cite{pohl2014dynamic}
interactions have also been identified as potentially arresting MIPS in
simulations. For
the bacteria \emph{Myxococcus xanthus}, it has been argued that MIPS
leads to the formation of fruiting bodies which coarsen up to a
certain size before other mechanisms become important
(Fig.~\ref{fig:manips}g)~\cite{liu_self-driven_2019}.  Furthermore, in the
bacteria that are genetically
engineered to experience quorum-sensing interactions~\cite{liu_sequential_2011}, MIPS is arrested to
a typical size because of the population dynamics of the
bacteria~\cite{cates_arrested_2010}, thereby forming the alternating
rings reproduced in Fig.~\ref{fig:manips}h. Finally, in other systems, attractive interactions are known to play an important role. Janus self-diffusiophoretic colloids have been reported to exhibit phase-separation or clustering far outside the MIPS region~\cite{theurkauff2012dynamic,palacci_living_2013,ginot_aggregation-fragmentation_2018}. Assessing the respective roles of MIPS and of attractive forces in these systems remains a challenge.

% Theory:
% \begin{itemize}
% \item QSPAPs (AOUPs, RTPs, ABPs, passive with $T(\rho)$
% \item PFAPs (AOUPs, RTPs, ABPs)
% \item Lattice models
% \end{itemize}
% Shall we discuss large-scale descriptions here ? Maybe not.

% Experiments:
% \begin{itemize}
% \item Self-propelled colloids (Cottin-Bizonne, Bechinger, Palacci)
% \item AC colloids (Dauchot)
% \item Quorum-sensing (Bechinger)
% \item Bacteria (JD, Shaevitz)
% \item In a polar fluid: Bartolo
% \end{itemize}

\section{Position-dependent propulsion speed and mean-field instability towards MIPS}
\label{sec:handwaving}
A first ingredient in the instability leading to MIPS is the
propensity of active particles to accumulate where they move
slower. To establish this fact, let us consider self-propelled
particles with an isotropic reorientation mechanism and a self
propulsion speed that is position dependent:
$\dot \bfr = v(\bfr) \bfu$. The master equation governing the
evolution of the probability to find a particle at position $\bfr$,
going in direction $\bfu$ reads
\begin{equation}\label{eq:mastervofr-theta}
    \partial_t P(\bfr,\bfu,t)=-\grad \cdot [ v(\bfr) \bfu P(\bfr,\bfu,t)] + \Theta P(\bfr,\bfu,t)
\end{equation}
where $\Theta$ is a linear operator accounting for the dynamics of $\bfu$. In 2D, the particle orientation can be parametrized as $\bfu=(\cos\theta,\sin\theta)$. ABPs then correspond to 
$\Theta P(\bfr,\theta)=D_r \partial_{\theta\theta}P(\bfr,\theta)$, while RTPs are described by
$\Theta P=-\alpha P(\bfr,\theta)+\int \frac{\alpha d \theta'}{2\pi}
P(\bfr,\theta')$. In both cases, the steady-state solution
of~\eqref{eq:mastervofr-theta} simply reads
\begin{equation}\label{eq:Pvfor}
    P(\bfr,\bfu)=\kappa/v(\bfr)\;,
\end{equation} 
where $\kappa$ is a constant ensuring
normalization. Active particles thus accumulate where they move
slower, with the $1/v(\bfr)$ scaling simply measuring the variation in
residence time when the propulsion speed is non-uniform. It is interesting to note that gradients of $v(\bfr)$, which locally break the isotropy
of space, do not lead to a locally anisotropic distribution of orientations.

\begin{figure}
  \includegraphics[width=\textwidth]{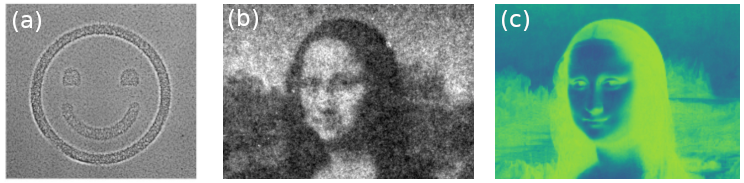}
  \caption{Experiments on light-activated bacteria where  Eq.~(\ref{eq:Pvfor}) is used to control the spatial organization of the bacteria. (a) The bacteria converge into a smiley pattern. (b) Pattern in bacterial density created after illuminating the bacteria with the light pattern shown in (c). Panel (a) is reproduced from Ref.~\cite{arlt_painting_2018} and panels (b,c) from Ref~\cite{frangipane_dynamic_2018}. All are under the \href{https://creativecommons.org/licenses/by/4.0/}{Creative Commons licence}.}\label{fig:ligh-bacteria}
\end{figure}

This simple result has been verified
experimentally and used to control the spatial organisation of
bacteria. Using bacteria whose swim speed can be controlled by light, the spatial modulation of the light field can be used to generate a   position-dependent $v(\bfr)$. In turn, this allows organizing the bacteria into
complex patterns~\cite{arlt_painting_2018,frangipane_dynamic_2018}, as shown in Fig.\ref{fig:ligh-bacteria}.

The instability leading to MIPS then emerges from the interplay between this tendency of
active particles to accumulate where they go slower and interactions
that slow down the particles at high density. The particles then tend to accumulate where they are already atypically dense, hence triggering phase
separation. Interestingly, this narrative can be turned into a
quantitative criterion to predict a linear instability.

\begin{figure}
  \centering
  \includegraphics[width=\textwidth]{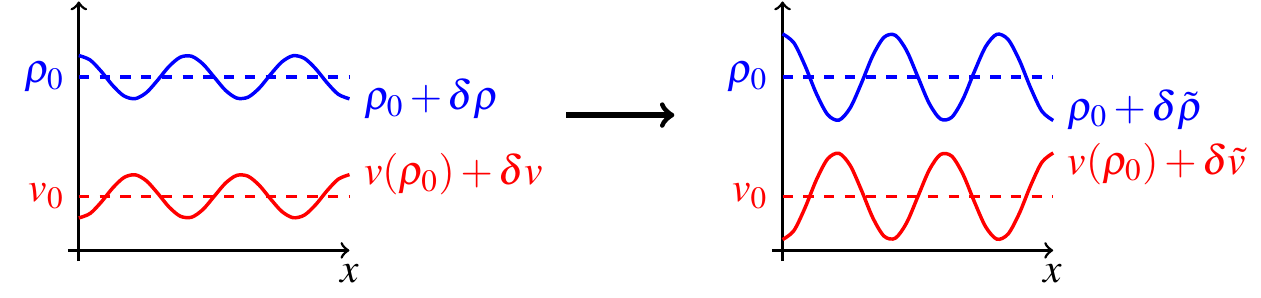}
  \if{
    \begin{tikzpicture}[domain=0:2.5,scale=1.1]
      \draw (2.5,0) node[below] {$x$};
      
      \draw[->,thick] (0,-0.1) -- (0,2.3);
      \draw[->,thick] (-0.1,0) -- (2.5,0);
      
      \draw[blue,dashed,line width=1pt] (0,1.6) -- (2.5,1.6);
      \draw[red,dashed,line width=1pt] (0,.5) -- (2.5,.5);

    \draw[blue] (0,1.6) node[anchor=east] {$\rho_0$};
    \draw[red] (0,.5) node[anchor=east] {$v_0$};
    \draw[blue,line width=1pt] plot[smooth] function{.2*cos(2*3.14*(x))+1.6};
    
    \draw[red,line width=1pt] plot[smooth] function{.2*cos(2*3.14*(x)+3.14)+.5};
    \draw[blue] (2.5,1.4) node[anchor=west] {$\rho_0+\delta\rho$};
    \draw[red] (2.5,.7) node[anchor=west] {$v(\rho_0)+\delta v$};
    
    \draw[ultra thick,->] (4.5,1.25) -- (5.5,1.25);
    
    \begin{scope}[xshift=6.5cm]
      \draw (2.5,0) node[below] {$x$};

      \draw[->,thick] (0,-0.1) -- (0,2.3);
      \draw[->,thick] (-0.1,0) -- (2.5,0);
      
      \draw[blue,dashed,line width=1pt] (0,1.6) -- (2.5,1.6);
      \draw[red,dashed,line width=1pt] (0,.5) -- (2.5,.5);

      \draw[blue] (0,1.6) node[anchor=east] {$\rho_0$};
      \draw[red] (0,.5) node[anchor=east] {$v_0$};
      \draw[blue,line width=1pt] plot[smooth] function{.4*cos(2*3.14*(x))+1.6};
      
      \draw[red,line width=1pt] plot[smooth] function{.4*cos(2*3.14*(x)+3.14)+.5};
      \draw[blue] (2.5,1.3) node[anchor=west] {$\rho_0+{\delta\tilde \rho}$};
      \draw[red] (2.5,.8) node[anchor=west] {$v(\rho_0)+{\delta \tilde v}$};
    \end{scope}
    
  \end{tikzpicture}}\fi
  \caption{Illustration of the linear instability leading to MIPS. An
    initial density fluctuation leads to a modulation in propulsion
    speed. If $v(\rho_0)$ decreases sufficiently fast with the density $\rho_0$,
    satisfying the criterion Eq.~(\ref{eq:criterion}), it leads in
    turn to amplifying the density fluctuation. The homogeneous system
    is then linearly unstable.}
  \label{fig:MIPSLS}
\end{figure}
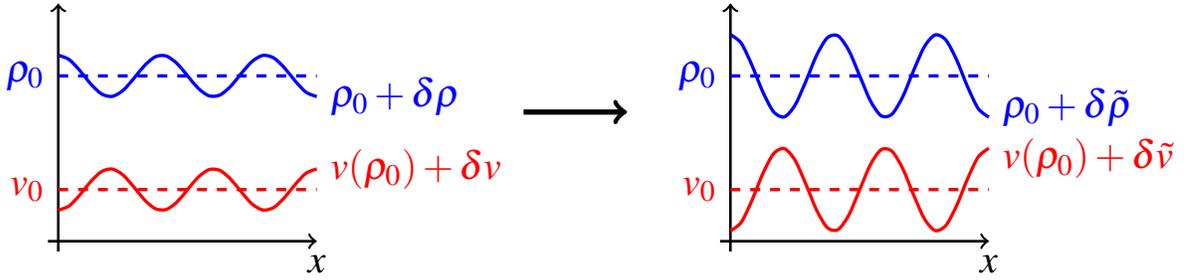

Consider QSAPs that move at a speed locally given by $v(\rho(\bfr))$. Let us study the fate of a small fluctuation of the density field, $\rho(\bfr)=\rho_0+\delta\rho(\bfr)$, around its average value $\rho_0$ (See
Fig.~\ref{fig:MIPSLS}). To linear order, the velocity is given by
\begin{equation}
  \label{eq:vrho-dvlpt}
    v(\rho(\bfr))=v(\rho_0) + \delta \rho(\bfr) v'(\rho_0)
\end{equation}
where $v'(\rho_0)\equiv dv/d\rho$ at $\rho=\rho_0$. As we have shown
above, the density field tends to relax toward
$\rho(\bfr)=\tilde\kappa/v(\rho(\bfr))$ where $\tilde\kappa$ is a normalization
constant. Using Eq.~(\ref{eq:vrho-dvlpt}), one finds, to first order in
$\delta \rho$,
\begin{equation}\label{eq:intermediaire1}
    \rho(\bfr)=\frac{\tilde\kappa}{v(\rho_0)}\left(1-\frac{v'(\rho_0)\delta \rho}{v(\rho_0)}\right)\;.
\end{equation}
Mass conservation imposes $\int d\bfr \delta\rho(\bfr)=0$ so that integrating   Eq.~\eqref{eq:intermediaire1} over space leads to $\tilde\kappa=\rho_0 v(\rho_0)$. The density field then  tends to relax towards a new perturbation $\rho(\bfr)\equiv \rho_0+\delta \tilde\rho$, with
\begin{equation}
 \delta \tilde\rho = -\rho_0\frac{v'(\rho_0)}{v(\rho_0)} \delta\rho
\end{equation}
This means that the initial perturbation is amplified whenever the
decrease of $v$ with $\rho_0$ is sufficiently strong:
\begin{equation}
  \label{eq:criterion}
    \frac{v'(\rho_0)}{v(\rho_0)}<-\frac{1}{\rho_0} \quad \text{i.e.} \quad \frac{d}{d\rho_0} [\rho_0 v(\rho_0)] <0\;.
\end{equation}

The instability criterion~\eqref{eq:criterion} will be derived more rigorously in Section~\ref{sec:generalized thermo}. However, this
heuristic argument already captures qualitatively the instability
mechanism leading to MIPS in most models. Although it is derived for
QSAPs moving with a speed $v(\rho(\bfr))$, it applies also to $N$ PFAPs
for which an effective propulsion speed can be defined as
$v_{\rm eff}\equiv N^{-1} \sum_i \langle \dot \bfr_i \cdot \bfu_i \rangle$. This effective speed, which is averaged over all the particles, measures how fast particles move along their orientations. It tends to decrease with
increasing density because of collisions. Simulation measurements show a linear decay $v_{\rm eff}(\rho_0)= v_0 - c \rho_0 $,
with a constant $c$, up to rather high packing
fractions~\cite{Fily2012PRL}, which indeed satisfy the instability
criterion Eq.~(\ref{eq:criterion}) as soon as $\rho_0>1/(2c)$.

\section{Large-scale long-time description in the presence of
position-dependent propulsion speed}
\label{sec:vofr} Before discussing in more details the case of interacting particles,
let us consider the long-time large-scale dynamics of $N$
non-interacting active particles with a position-dependent
self-propulsion speed $v(\bfr)$ that varies slowly in space: in a
system of size $L$, we consider $|\grad v|\sim {\cal O}(L^{-1})$.

\subsection{A Fokker-Planck approximation}
\label{sec:vofrFP} For the sake of clarity, we start from the case of
a single particle, work in two space dimensions, and consider only
tumbles with rate $\alpha$. The generalization to rotational diffusion
and higher dimensions is straightforward~\cite{Cates:2013:EPL}. We
also account for the presence of Brownian noise leading to a
translational diffusivity $D_t$. The master equation describing the
dynamics is then given by
\begin{equation}\label{eq:mastervofr} \partial_t
P(\bfr,\theta,t)=-\grad \cdot [v(\bfr) \bfu(\theta) P(\bfr,\theta,t) -
D_t \grad P(\bfr,\theta,t)] - \alpha
P(\bfr,\theta,t)+\frac{\alpha}{2\pi} p(\bfr,t)\;
\end{equation} where we have introduced the probability density to
find the particle at position $\bfr$, irrespective of its direction,
\begin{equation} p(\bfr,t)=\int d\theta P(\bfr,\theta,t)
\end{equation}

Let us now construct the long-time large-scale dynamics of
$p(\bfr,t)$. Integrating Eq.~\eqref{eq:mastervofr} over $\theta$ leads
to
\begin{equation}\label{eq:dynpsi} \partial_t p(\bfr,t)=-\grad \cdot
[v(\bfr) \bfm (\bfr,t)] + D_t \grad^2 p(\bfr,t)\quad\text{where}\quad
\bfm(\bfr,t) \equiv \int d\theta \bfu(\theta)P(\bfr,\theta,t)\;.
\end{equation} The field $\bfm(\bfr,t)$ measures the average
orientation of the particle at position $\bfr$, whereas $v(\bfr) \bfm
(\bfr,t)$ is the contribution to the current in position space due to
self-propulsion. To close the dynamics of $p(\bfr,t)$, it is useful to
compute that of $\bfm(\bfr,t)$ by multiplying
Eq.~\eqref{eq:mastervofr} by $\bfu(\theta)$ and integrating over
$\theta$. Component per component, this leads to
\begin{equation} \partial_t m_i(\bfr,t)=-\partial_j
\cdot \Big[v(\bfr) \int d\theta u_i u_j P (\bfr,\theta,t)\Big] + D_t
\grad^2 m_i(\bfr,t)-\alpha m_i(\bfr,t)\;,
\end{equation} where $\partial_j\equiv\frac{\partial}{\partial x_j}$
is a short-hand notation for the $j^{th}$ component of the gradient
operator. In all the Chapter, summation over repeated indices is implied. Introducing the tensor $Q_{ij}(\bfr)=\int d\theta
\big[u_iu_j-\frac{\delta_{ij}}{2}\big]P(\bfr,\theta)$, which measures
the average local nematic order, the dynamics of $\bfm(\bfr,t)$ can
then be written as
\begin{equation}\label{eq:dynm} \partial_t \bfm (\bfr,t)=-\grad
\Big[\frac{v(\bfr)}2 p\Big] - \grad \cdot [v(\bfr) \mathbf{Q}(\bfr)] + D_t \grad^2
\bfm(\bfr,t)-\alpha \bfm(\bfr,t)\;.
\end{equation} Finally, multiplying Eq.~\eqref{eq:mastervofr} by
$u_{i}u_j -\frac{\delta_{ij}}2$ and integrating over $\theta$ shows
the dynamics of $Q_{ij}$ to take the form
\begin{equation}\label{eq:dynQ} \partial_t Q_{ij} (\bfr,t)=-\grad
\cdot [\dots] -\alpha Q_{ij}(\bfr,t)\;,
\end{equation} where the precise form of the gradient term will be shown to be
irrelevant. The Equations~\eqref{eq:dynpsi},~\eqref{eq:dynm}
and~\eqref{eq:dynQ} are the three first equations of an infinite
hierarchy that couples the dynamics of the various moments of
$P(\bfr,\theta)$. Let us now show that, in the long-time large-scale
limit, we can close this hierarchy.

In the presence of a field $v(\bfr)$ varying smoothly over a large
scale $L$, the steady-state solution in the absence of translational
diffusivity satisfies $P(\bfr,\theta) \propto 1/v(\bfr)$, and it
 thus reproduces the smooth large-scale variations of $v(\bfr)$ as
long as the latter remain bounded away from zero. The Brownian noise
is only expected to alter quantitatively this behaviour and we thus
expect the late-time dynamics of $p(\bfr)$ to describe the relaxation
of large wave-length modes. Since $p(\bfr)$ is a conserved field, the
time it takes to relax over a length scale $L$ diverges as $L\to
\infty$: $p(\bfr)$ is a slow, hydrodynamic mode. On the contrary, the
dynamics~\eqref{eq:dynm} and~\eqref{eq:dynQ} show the relaxation times
of $\bfm(\bfr)$ and $\mathbf{Q}(\bfr)$ to scale as $\alpha^{-1}$, which remains
finite as $L$ diverges. These fast modes are thus enslaved to the sole
slow mode $p(\bfr,t)$: $\bfm(\bfr,t)$ and $Q_{ij}(\bfr,t)$ follow quasi-statically the
solutions of Eq.~\eqref{eq:dynm} and~\eqref{eq:dynQ} in which we set
$\partial_t \bfm=\partial_t Q_{ij}=0$. Furthermore, in the limit of
large $L$, $|\grad v|\sim {\cal O}(L^{-1})$ so that all the fields are
expected to vary slowly in space and, as we now show, we can use a
gradient expansion to close the dynamics for $p(\bfr)$.  Inspection of
Eq.~\eqref{eq:dynQ} shows that, to leading order, $Q_{ij}\sim {\cal
O} (\grad) \sim {\cal
O}(L^{-1})$. In turns, this leads to
\begin{equation}\label{eq:mslaved} \bfm(\bfr,t) \simeq -\frac 1 \alpha
\grad \left[\frac{v(\bfr)}{2} p(\bfr)\right] +{\cal O}(\grad^2)
\end{equation} Inserting Eq.~\eqref{eq:mslaved} into
Eq.~\eqref{eq:dynpsi} then leads, to leading order, to the following
dynamics for $p(\bfr)$:
\begin{equation}~\label{eq:FP1} \partial_t p(\bfr,t)=\grad \cdot
\left[\frac{v(\bfr)}{2\alpha} \grad [v(\bfr) p(\bfr,t)] + D_t \grad
p(\bfr,t)\right]\;.
\end{equation}

A first interesting result that can be obtained from
Eq.~\eqref{eq:FP1} is that the flux-free steady-state distribution
satisfies
\begin{equation} \frac{v(\bfr)\grad [v(\bfr)]}{2\alpha} p(\bfr,t) =
-\left(\frac{v^2(\bfr)}{2\alpha} + D_t\right) \grad
p(\bfr)\quad\text{so that}\quad \grad \ln p(\bfr)=-\frac 1 2 \grad \ln
\left(\frac{v^2(\bfr)}{2\alpha} + D_t\right)
\end{equation} To leading order in gradients, the steady-state
distribution is thus given by
\begin{equation}\label{eq:steadystateDt} p(\bfr)\propto
\frac{1}{\sqrt{\frac{v^2(\bfr)}{2\alpha} + D_t}}\;,
\end{equation} where, again, we disregard normalization issues. This
result generalizes Eq.~\eqref{eq:Pvfor} to finite translational
diffusion in the presence of smooth large-scale variations of
$v(\bfr)$.  Introducing $D_0=v^2/(2\alpha)$, which is the large-scale effective diffusivity
of an RTP of self-propulsion speed $v$ in 2d, the steady-state distribution $p(\bfr)$ is then found to be, to first order in $D_t$, $p
\propto v^{-1} (1 - \frac{D_t}{2 D_0})$. As expected, $D_t$ can be
neglected when it is much smaller than $D_0$.

Reproducing the computations above in $d$ space dimensions and
including rotational diffusion as well as tumbles leads to a
late-time dynamics~\cite{Cates:2013:EPL}:
\begin{equation}\label{eq:FP} \partial_t p(\bfr,t) = \grad \cdot
\big[ \grad [D_c(\bfr) p(\bfr,t)] - \bfF(\bfr) p(\bfr,t)\big]
\end{equation} where
\begin{equation}\label{eq:DandF} D_c(\bfr)=D_t+\frac{v^2(\bfr)
\tau}d\quad\text{and}\quad \bfF({\bfr})=\frac 1 2 \grad
\left[D_t+\frac{v^2(\bfr)\tau}d\right]
\end{equation} and we have introduced the persistence
time of the active particle
\begin{equation} 
\tau\equiv[\alpha + (d-1) D_r]^{-1}\;.
\end{equation} 
The dynamics~\eqref{eq:FP} is
nothing but a Fokker-Planck approximation to the active dynamics in
the presence of a non-uniform self-propulsion speed. Having integrated out the orientational degrees of freedom of the active particles, the long-time
large-scale dynamics of their positions are thus equivalent to an
It\=o-Langevin dynamics:
\begin{equation}\label{dyn:Itovofr} \dot {\bfr}=\bfF(\bfr)+\sqrt{2
D_c} \boldsymbol\eta\;,
\end{equation} where $\boldsymbol\eta$ is a $d$-dimensional Gaussian
white noise whose spatial components satisfy $\langle
\eta_{\mu}\eta_{\nu}\rangle=\delta_{\mu\nu} \delta(t-t')$.

A number of comments are in order. First, the steady-state
solution~\eqref{eq:steadystateDt} is a flux-free solution of the Fokker-Planck equation~\eqref{eq:FP}. It thus corresponds to an equilibrium state: the
microscopic, non-equilibrium nature of the active dynamics has
disappeared upon coarse-graining. Then, inspection of the
It\=o-Langevin dynamics~\eqref{dyn:Itovofr} shows that $\bfF$ and
$\sqrt{2 D_c}\eta_i$ favors opposite trends for the particles:
$\bfF$ drives particles towards high-speed regions whereas $\sqrt{2
D_c}\eta_i$ drives them towards low-activity ones. Finally, note that
Eq.~\eqref{eq:FP} also corresponds to a Stratonovich-Langevin
dynamics, which would also be given by Eq.~\eqref{dyn:Itovofr}, albeit
with $\bfF=0$. In the next section, we will use It\=o calculus to
construct the fluctuating hydrodynamics describing $N$ non-interacting
active particles and we will thus stick to the It\=o prescription.

\subsection{Fluctuating hydrodynamics}\label{sec:flucthydrvofr} To keep notations as light as
possible, we first work in one dimension before generalizing our
results to higher dimensions.  We consider $N$ active particles at
positions $x_i$, evolving in a field $v(x)$, in the limit in which the
particle dynamics are well approximated by
Eq.~\eqref{dyn:Itovofr}. The purpose of this section is to describe
the stochastic evolution of the density field
\begin{equation}\label{eq:rho} \hat \rho(x,t)=\sum_{i=1}^N
\delta[x-x_i(t)] \;.
\end{equation} Note that the time dependency of $\hat \rho(x,t)$
solely stems from that of $x_i(t)$. Constructing the dynamics of the
density field thus amounts to applying the chain rule to
Eq.~\eqref{eq:rho}. This is, however, a non-trivial task because
$x_i(t)$ is far from being a smooth function of time. The standard
chain rule, built using differential calculus, fails for stochastic
dynamics like Eq.~\eqref{dyn:Itovofr}. Instead, a proper
regularization is needed, using stochastic calculus, which leads to
the so-called It\=o-formula. A rigorous derivation of the latter can
be found in textbooks on stochastic
calculus\cite{oksendal2003stochastic} and we instead present here a
heuristic approach.

To determine $\frac{d}{dt}\hat \rho(x,t)$, we need to estimate $\Delta
\rho_i(t)\equiv \delta[x-x_i(t+dt)]-\delta[x-x_i(t)]$ to first order
in $dt$. Taylor-expanding $\Delta \rho_i$ to second order yields
\begin{equation}\label{eq:Taylor} \frac{\Delta \rho_i(t)}{dt}=-
\frac{x_i(t+dt)-x_i(t)}{dt} \frac{\partial}{\partial x}
\delta[x-x_i(t)]+\frac 1 2 \frac{[x_i(t+dt)-x_i(t)]^2}{dt}
\frac{\partial^2}{\partial x^2} \delta[x-x_i(t)]\;.
\end{equation} The first term on the right-hand side yields $-\dot
x_i\delta'[x-x_i(t)]$ when $dt$ is sent to zero. In the usual chain
rule, this is the sole contribution to $\partial_t
\delta[x-x_i(t)]$ and the second term in the right-hand side of
Eq.~\eqref{eq:Taylor} can be neglected. Let us now show that this is
not the case here. To leading order in $dt$, integrating Eq.~\eqref{dyn:Itovofr} in 1d between
$t$ and $t+dt$ leads to
\begin{equation}
  \begin{aligned} \Delta x_i(t)\equiv x_i(t+dt)-x_i(t)&=\int_t^{t+dt}
\!\!\!\!\!\!\!\!ds F[x_i(s)]+ \int_t^{t+dt} \!\!\!\!\!\!\!\!ds \sqrt{2
D_c[x_i(s)]} \eta_i(s)\\ &\simeq F[x_i(t)] dt + \sqrt{2 D_c[x_i(t)]}
d\eta_i(t) \;,\label{eq:approxdx}
  \end{aligned}
\end{equation} where we have introduced $d\eta_i(t)\equiv
\int_t^{t+dt} ds \eta_i(s)$.  Taking the square of
Eq.\eqref{eq:approxdx} then leads to $\Delta x_i(t)^2= 2 D_c[x_i(t)] d
\eta_i(t)^2 + o(dt)$, where we have used that $d\eta_i(t)$ vanishes as
$dt$ goes to zero. The second term in the right-hand side of
Eq.~\eqref{eq:Taylor} will thus contribute if $d \eta_i(t)^2$ is of
order $dt$. To show that this is indeed the case, we compute the
successive moments of $d\eta_i(t)^2$:
\begin{equation*} \langle d\eta_i(t)^2\rangle=
\int_t^{t+dt}\!\!\!\!\!\!\!\!\!\! du \int_t^{t+dt}\!\!\!\!\!\!\!\!\!\!
ds \langle \eta_i(s)\eta_i(u)\rangle = \int_t^{t+dt}
\!\!\!\!\!\!\!\!\!\! du \int_t^{t+dt} \!\!\!\!\!\!\!\!\!\! ds
\delta(s-u) = dt
\end{equation*} The average of the random variable $d\eta_i(t)^2$ is
thus equal to $dt$. The higher moments of $d\eta_i(t)^2$, which
characterize its fluctuations, can be computed using Wick theorem;
they are negligible to order $dt$. For instance, $\langle
d\eta_i(t)^4\rangle=6 D_c[x_i(t)]^2 dt^2$. It is thus tempting to
simply replace $[x_i(t+dt)]-x_i(t)]^2$ by its average value $2
D_c[x_i(t)] dt$ in Eq.~\eqref{eq:Taylor}. This would lead to:
\begin{equation}\label{eq:Itorho_i} \partial_t \delta[x-x_i(t)]=- \dot
x_i(t) \partial_x \delta[x-x_i(t)] + D[x_i(t)] \partial_{xx}
\delta[x-x_i(t)]\;,
\end{equation} which is indeed the result of applying It\=o formula to
$\delta[x-x_i(t)]$.  Note that a rigorous construction of It\=o
formula is significantly more involved than the hand-waving argument
above and it involves clarifying the meaning of the equality
in~\eqref{eq:Itorho_i}\cite{oksendal2003stochastic}.

Using that, for all functions $f$,
\begin{equation} f(x_i) \partial_x \delta(x-x_i)= \partial_x[ f(x_i)
\delta(x-x_i)]= \partial_x [f(x) \delta(x-x_i)]\;,
\end{equation} Eq.~\eqref{eq:Itorho_i} can be rewritten as
\begin{equation}\label{eq:Itorho_ii} \partial_t \delta[x-x_i(t)]=-
\partial_x\Big[ F(x) \delta[x-x_i(t)] + \sqrt{2 D[x(t)]} \eta_i
\delta(x-x_i) - \partial_{x} \big[D[x(t)]\delta[x-x_i(t)]\big] \Big]
\end{equation} Summing upon $i$ then leads to
\begin{equation}\label{eq:Itorho_iii} \partial_t \rho(x,t)=-
\partial_x\big[ F(x) \rho(x,t) + \sqrt{2 D[x(t)] \rho(x,t)}
\Lambda(x,t) - \partial_{x} [D[x(t)]\rho(x,t) \big]\big]\;,
\end{equation} where $\Lambda(x,t)$ is a Gaussian noise field of zero
mean and unit variance and we used that~\cite{dean1996langevin}:
\begin{equation} \langle \sum_{i,j} \delta[x-x_i(t)] \eta_i(t)
\delta[x'-x_j(t')] \eta_j(t') \rangle = \rho(x,t) \delta(t-t')
\delta(x-x')\;.
\end{equation}

The above derivation can be reproduced in higher dimensions. Using the
expressions~\eqref{eq:DandF} for $D_c$ and $F$, it leads to
\begin{equation}\label{eq:coldynr} \partial_t
\rho(\bfr,t)=\grad\cdot\left[M \grad \mu(\bfr,[\rho])+\sqrt{2 M}
\boldsymbol{\Lambda}(\bfr,t)\right]
\end{equation} where $\mu$ acts as a chemical potential given by
\begin{equation}\label{eq:colmur} \mu(\bfr,[\rho])=\ln\left[\rho(\bfr)
\sqrt{D_c(\bfr)}\right]
\end{equation} and the collective mobility $M$ reads
\begin{equation}\label{eq:colmob} M(\bfr)=\rho(\bfr)
D_c(\bfr)\qquad\text{with}\qquad D_c(\bfr)=D_t+\frac{v^2(\bfr)\tau}d.
\end{equation}

Equation~\eqref{dyn:Itovofr} is a diffusive approximation, at the
single-particle level, of a run-and-tumble dynamics with
position-dependent self-propulsion speed $v(\bfr)$. In
Section~\ref{sec:vofrFP}, we have shown that this approximate dynamics
satisfies detailed balance and thus amounts to an equilibrium
dynamics. Equations~\eqref{eq:coldynr},~\eqref{eq:colmur}
and~\eqref{eq:colmob} are its collective counterparts. Naturally, they
also correspond to an equilibrium dynamics and the chemical potential
$\mu$ is the functional derivative of a free energy:
\begin{equation}\label{eq:FEvofr} \mu(\bfr,[\rho])=\frac{\delta
\mathcal{F}[\rho]}{\delta \rho(\bfr)}\qquad\text{with}\qquad
\mathcal{F}[\rho]=\int d\bfr \Big[\rho(\bfr)[\ln\rho(\bfr)-1] +\frac 1 2
\rho(\bfr)\ln[D_c(\bfr)]\Big]
\end{equation} The first term in the effective free energy
$\mathcal{F}$ is an entropic term, which conveys that the persistent random walks
of run-and-tumble particles favor a uniform spreading of the
particles in space. The second term stems from the tendency of active
particles to accumulate where they go slower. Here, this takes the form
of an effective potential energy
$U(\bfr)=\ln[D_c(\bfr)]/2$. Equation~\eqref{eq:FEvofr} thus reveals an
intriguing mapping between active particles whose self-propulsion is
non-uniform and passive particles in potential energy landscapes. As
we shall show in the next section, this analogy partially extends to
interacting dynamics.

\section{Fluctuating hydrodynamics of Active Particles interacting via
Quorum-Sensing (QSAPs)}
\label{sec:QSAPs}

Quorum-sensing interactions are common in the biological world where
they describe situations in which the behaviours of individuals are
affected by the local density of their surrounding peers. A pedestrian
adapting its walking speed depending on the density of the surrounding crowd
is a simple macroscopic example of such interactions. At the
microscopic scale, quorum-sensing interactions are frequent among
cells, where they are typically mediated by signaling molecules. In
a bacterial population, a common mechanism is the production of
auto-inducers, like acylated homoserine lactones, which then
diffuse in the surrounding medium and regulate the gene expressions of
other bacteria. In this section, we retain this framework to make
contact with recent experiments on bacterial colonies in which the
cell motility is regulated by
quorum-sensing~\cite{liu_sequential_2011,curatolo2020cooperative}. To
lighten the notation, we neglect $D_t$, which is typically three
orders of magnitudes smaller that $D_0$ for swimming
\textit{E. coli}~\cite{berg2008coli,wilson2011differential}.

We note $c(\bfr)$ the concentration field of signaling molecules
produced by the bacteria at rate $\beta$. After their production,
these molecules diffuse in the surrounding medium with a diffusivity
$D$ and they are degraded with a rate $\gamma$, leading to the dynamics
\begin{equation}\label{eq:dync} \dot c(\bfr,t) =D \Delta c(\bfr,t) -
\gamma c(\bfr,t) + \beta \rho(\bfr,t)
\end{equation}

We aim at describing the large-scale collective dynamics of our
bacterial assembly on time scales where the division and death of
bacteria can be neglected, so that the bacterial density field
satisfies a conservation law $\partial_t \rho(\bfr,t)=-\nabla \cdot {\bf
J}(\bfr,[\rho])$, where `$[\rho]$' refers to a functional dependence of the current ${\bf J}$ on the density field $\rho(\bfr,t)$. The relaxation of a density modulation $\delta
\rho(x)$ around a homogeneous profile $\rho_0$ then satisfies
\begin{equation} \partial_t {\delta \rho(\bfr,t)}=-\nabla \cdot \int d\bfr '
\frac{\delta {\bf J}(\bfr,[\rho_0])}{\delta \rho(\bfr')}\delta
\rho(\bfr',t)\;.
\end{equation} When the system is invariant under translation,
$\frac{\delta {\bf J}(\bfr,[\rho_0])}{\delta \rho(\bfr')}\equiv {\bf
K}(\bfr-\bfr',\rho_0)$ only depends on $\bfr-\bfr'$ so that, in
Fourier space, the dynamics of the Fourier modes decouple:
\begin{equation}\label{eq:deltaqrelax}
\partial_t{\delta\rho({\bf q},t)}=i {\bf q} \cdot {\bf K}({\bf
q},\rho_0) \delta \rho({\bf q},t)\;.
\end{equation}
The typical relaxation time of the Fourier mode
$\delta \rho({\bf q},t)$ thus diverges as $[{\bf q} \cdot {\bf K}({\bf
q},\rho_0)]^{-1}$ as $q\to 0$: large-scale modes take a macroscopic time
to relax since matter has to be transported over a macroscopic size, which necessarily takes a macroscopic time at finite speed. Equation~\eqref{eq:deltaqrelax} is the generic reason why a hydrodynamic field can be associated to any conservation law. Here, mass conservation makes the density field a slow field.

On the contrary, the
dynamics~\eqref{eq:dync} leads to a finite relaxation time $\tau_q\sim
\gamma^{-1}$ as $q\to 0$ and we can thus consider that $c(\bfr,t)$
follows adiabatically the solution of
\begin{equation}\label{eq:screenedPoisson} (\gamma-D\Delta) c({\bfr,t)
= \beta \rho(\bfr,t)}
\end{equation} which is nothing but a screened Poisson dynamics for
the field $c(\bfr,t)$ in the presence of a source field
proportional to $\rho(\bfr,t)$. Equation~\eqref{eq:screenedPoisson} can be solved as
\begin{equation}\label{eq:solc} c(\bfr,t)=\int d\bfr' G(\bfr-\bfr')
\rho (\bfr',t)\;,
\end{equation} where $G$ is the corresponding Green's
function~\cite{jackson_classical_1999}, which satisfies
\begin{equation}\label{eq:defG}
    \left(\frac\gamma\beta - \frac D \beta \Delta  \right) G(\bfr)=\delta(\bfr) \;.
\end{equation} 
The concentration of signaling molecules thus acts as a slightly non-local proxy for the bacterial concentration through Eq.~\eqref{eq:solc}.

In turn, bacteria adapt their swimming patterns to the local
concentration $c(\bfr,t)$. Once the concentration field has been
eliminated using Eq.~\eqref{eq:solc}, the microscopic parameters of
the run-and-tumble dynamics, like the swim speed $v$ or the tumbling
rate $\alpha$, thus depend on $\rho$ in a slightly non-local way. Note
that quorum-sensing in bacterial suspensions involves a delay between
the variation of the level of a signaling molecules $c(\bfr,t)$ and
the change of the expression level of the proteins under the control
of the auto-inducer. In turn, this implies a delay between the
variation of $c(\bfr,t)$ and the adaptation of the swimming pattern of
the bacteria. We neglect this aspect here, assuming again that $c(\bfr,t)$
evolves slowly, following $\rho(\bfr,t)$ adiabatically. The coarse-grained
dynamics of run-and-tumble bacteria whose tumbling rate, duration and
swimming speed depend on the local density can be computed using the
methods described in this
chapter~\cite{Cates:2013:EPL,solon_active_2015,curatolo2020cooperative}. For simplicity, we discuss here only the case in which the velocity is
influenced by the local density.

Finally, the approximation of the run-and-tumble dynamics by a
diffusive one takes place on a time scale that is large compared to
the tumbling rates and durations, but which is small compared to that
governing the large-scale dynamics of the density field. We can thus treat
a self-propulsion speed $v(\bfr,[\rho])$ as simply depending on $\bfr$
to coarse-grain the bacterial dynamics into its diffusion-drift approximation~\eqref{dyn:Itovofr} and only let $\rho(\bf r,t)$ evolve on
much larger time scales. The result of this adiabatic treatment is to
predict an evolution of the density field given by
\begin{equation}\label{eq:coldyn} \partial_t
\rho(\bfr,t)=\grad\cdot\left[M \grad \mu(\bfr,[\rho])\right]+\sqrt{2 M}
\boldsymbol{\Lambda}(\bfr,t)
\end{equation} where $\mu$ and $M$ are now given by
\begin{equation}\label{eq:colmu} \mu(\bfr,[\rho])=\ln\left[\rho(\bfr)
v(\bfr,[\rho])\right]\qquad\text{and}\qquad M=\rho(\bfr) \frac{v^2(\bfr,[\rho])\tau}d
\;.
\end{equation} Note that going from the Langevin dynamics~\eqref{dyn:Itovofr} to the stochastic field dynamics~\eqref{eq:coldyn} when the velocity is a functional of the density field is
actually trickier than what was presented in
Section~\ref{sec:flucthydrvofr} because of the non-local dependency of
$M$ on the field $\rho$. The corresponding multiplicative noise has to
be handled with care and, in practice, Eq.~\eqref{eq:coldyn} is valid
for kernels $G$ which are symmetric~\cite{solon_active_2015}.

We now turn to the analysis of the dynamics~\eqref{eq:coldyn} to
characterize the motility-induced phase separation that can emerge
from quorum-sensing interactions.

\section{Generalized thermodynamics of QSAPs}
\label{sec:generalized thermo}
In general, the stochastic field theory~\eqref{eq:coldyn} does not describe an equilibrium dynamics and its steady-state distribution $P[\rho]$ is unknown. However, we show in this section that there are a number of cases in which connection with equilibrium physics can be made. 

Let us first note that the noise amplitude and the mobility are both given by $M(\bfr,[\rho])$ and thus satisfy a Stokes-Einstein relation. At this level of description, it is thus not the noise entering the field theory that drives the system out of equilibrium. 

The same cannot be said, however, for the generalized chemical potential $\mu(\bfr,[\rho])$. The latter induces an equilibrium dynamics if and only if it can be written as the functional derivative of a free energy, {\it i.e.} iff there is a functional ${\cal F}[\rho]$ such that 
\begin{equation}\label{eq:integ}
    \mu(\bfr,[\rho])=\frac{\delta \mathcal{F}[\rho]}{\delta \rho(\bfr)}\;.
\end{equation}
Equation~\eqref{eq:integ} need not admit a solution for arbitrary functionals $\mu(\bfr,[\rho])$. In finite dimensions, the Poincare lemma tells us that a vector field ${\bfF}$ is a gradient if its cross derivatives are equal\footnote{provided the space is simply connected}: $\partial_i F_j=\partial_j F_i$. This can be generalized to our field theory and Eq.~\eqref{eq:integ} admits a solution if and only if~\cite{o2020lamellar}:
\begin{equation}\label{eq:D}
    \mathcal{D}(\bfr,\bfr')= \frac{\delta \mu(\bfr,[\rho])} {\delta \rho(\bfr')}-\frac{\delta \mu(\bfr',[\rho])}{\delta \rho(\bfr)}=0\;.
\end{equation}
Note that Eq.~\eqref{eq:D} has to be understood in the sense of distributions, i.e. for any functions $\phi(\bfr)$, $\psi(\bfr')$, one needs $\int d{\bfr}d{\bfr'} \mathcal{D}(\bfr,\bfr')\phi(\bfr)\psi(\bfr')=0$. 

Consider, as an example, the case of $\mu(x,[\rho])= \partial_x^k \rho(x)$ in one dimension. Then, $\mathcal{D}(x,x')=(\partial_x^k-\partial_{x'}^k)\delta(x-x')$ and the integrability condition~\eqref{eq:D} reads
\begin{equation}
    0=\int d{x} [\phi^{(k)}(x)\psi(x)-\phi(x)\psi^{(k)}(x)]=    \int d{x} \phi^{(k)}(x)\psi(x) [1-(-1)^k]
\end{equation}
where the second equality stems from integrating by parts $k$ times. This condition holds for any pair of functions $(\psi,\phi)$ if and only if $k$ is an even number. In this case, the free energy is given by ${\cal F}=\frac 1 2 \int dx (-1)^n (\partial^n_x \rho)^2$, where $n=k/2$. Let us now study the conditions under which the large-scale dynamics of run-and-tumble particles interacting via quorum sensing admit a coarse-grained equilibrium description.

\subsection{The local approximation: an equilibrium theory}
\label{sec:local}
The self-propulsion speed $v(\bfr,[\rho])$ depends on the density through the concentration $c(\bfr,t)$ and can thus be written as
\begin{equation}\label{eq:vconvol}
    v(\bfr,[\rho])=v[\tilde \rho(\bfr)]\quad\text{with}\quad\tilde\rho(\bfr)=\int d\bfr' G(\bfr-\bfr') \rho(\bfr')\;.
\end{equation}
In Eq.~\eqref{eq:vconvol}, we have absorbed the normalization of $G$ in the definition of the function $v$ and set $\int d\bfr G(\bfr)=1$, which amounts to multiplying the left-hand side of Eq.~\eqref{eq:defG} by $\beta/\gamma$. $\tilde \rho(\bfr)$ is then the coarse-grained density measured at position $\bfr$ by the particles.
Note that, because the diffusion of $c$ is isotropic, so is $G$, leading to $\int d\bfr\, \bfr G(\bfr)=0$ and $\int d\bfr\, r_i\, r_j G(\bfr)=\delta_{i,j}  \ell_g^2$. Using Eq.~\eqref{eq:defG}, one finds---up to the normalization  mentioned above---that $\ell_g=\sqrt{2D/\gamma}$. The latter is the typical length over which the coarse-graining kernel $G$ samples the density field; it corresponds to the typical length travelled by signaling molecules before they are degraded. When the density field varies over much larger length scales, it is natural to approximate $    \tilde \rho(\bfr)$ as
\begin{equation}
\tilde \rho(\bfr)\simeq \int d\bfr' G(\bfr') [\rho(\bfr)-\bfr' \cdot \nabla \rho(\bfr) + \frac 1 2 \bfr'\cdot {\bf H}(\bfr) \bfr'] \;, 
\end{equation}
where $H_{ij}=\partial^2_{i,j} \rho(\bfr)$. Using the symmetry properties of $G$, this reduces to
\begin{equation}\label{eq:tilderho}
\tilde \rho(\bfr)\simeq \rho(\bfr)+ \frac {1} 2 \ell_g^2 \Delta \rho(\bfr)\;. 
\end{equation}

To leading order, $\tilde \rho(\bfr)=\rho(\bfr)$ and $\mu(r,[\rho])$ is a local function of $\rho(\bfr)$. In this local approximation, the integrability of $\mu(\bfr)=\ln[\rho(\bfr) v(\rho(\bfr))]$ is easily established since:
\begin{equation}
\begin{aligned}
    \mathcal{D}(\bfr,\bfr')&=\left(\frac{1}{\rho(\bfr)}+\frac {v'(\rho(\bfr))}{v(\rho(\bfr))}\right)\delta(\bfr-\bfr')-\left(\frac{1}{\rho(\bfr')}+\frac {v'(\rho(\bfr'))}{v(\rho(\bfr'))}\right)\delta(\bfr'-\bfr)=0\;,
    \end{aligned}
\end{equation}
where we have introduced $v'(\rho)\equiv \frac{d v(\rho)}{d\rho}$. The free energy ${\cal F}$ is then given by
\begin{equation}\label{eq:FEloc}
    \mathcal{F}[\rho]=\int d\bfr f(\rho(\bfr))\;,
\end{equation}
where $f(\rho)$ is the free energy density, given by
\begin{equation}\label{eq:FEDloc}
    f(\rho)=\rho(\ln \rho-1) +\int^\rho ds \ln v(s)\;.
  \end{equation}
  (Note that the integral is taken with respect to the argument of
  $v(\rho)$). Some comments are in order. First, the addition of a
  linear term $\mu_0 \rho$ to $f(\rho)$ simply leads to the constant
  shift
\begin{equation}
\mathcal{F} \to \mathcal {F}+\mu_0 \int d\bfr \rho(\bfr) =  \mathcal {F}+\mu_0 N\;,
\end{equation}
where $N$ is the total number of particles. Since the latter is conserved, this shift does not change the relative weights of different density profiles. This thus means that rescaling the velocity $v(s)\to z_0 v(s)$, which induces such a shift with $\mu_0=\ln z_0$, should not change the steady-state distribution. Here, nothing competes with $v(\rho(\bfr))$ to set a scale above which MIPS can be observed. It would not be the case if we had retained a finite translational diffusion $D_t$. Then, there would be a minimal self-propulsion speed above which MIPS can be observed ~\cite{cates2015motility}.

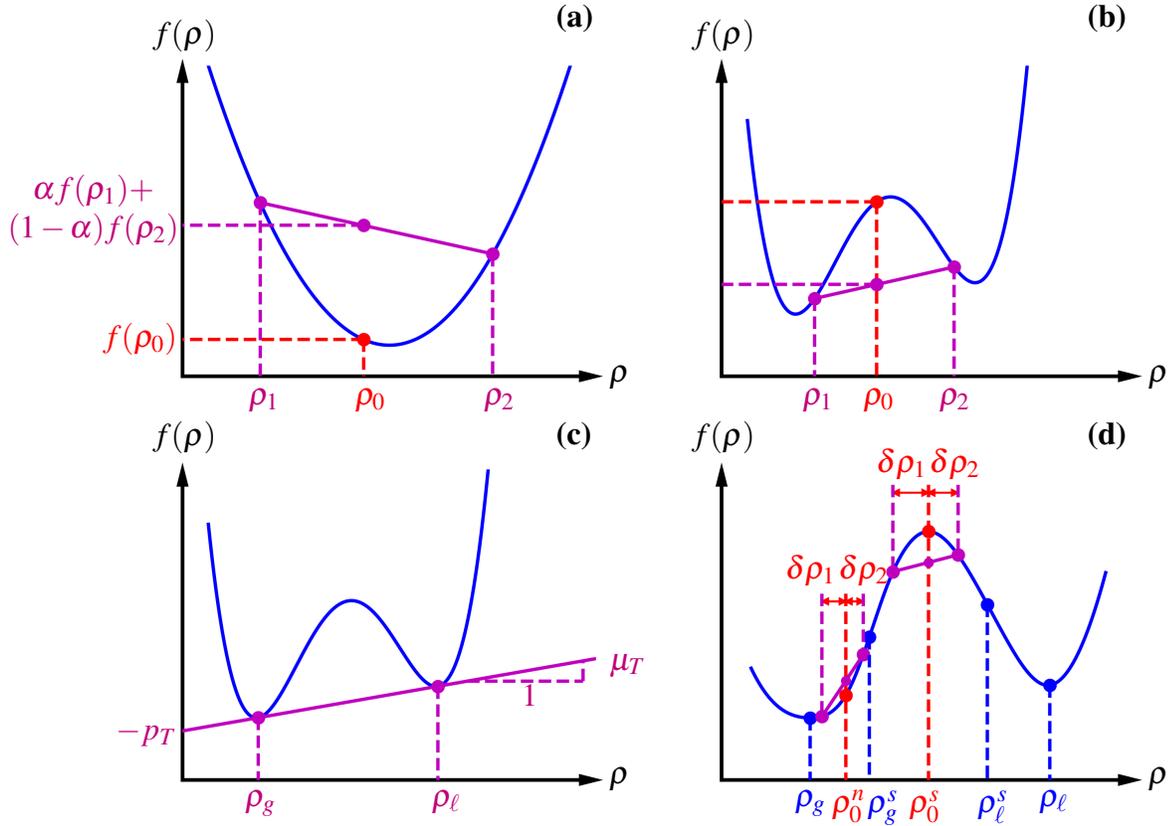
\begin{figure}
    \centering
    \if{
    \begin{tikzpicture}
    \path (0,0) node {\includegraphics[width=.8\textwidth]{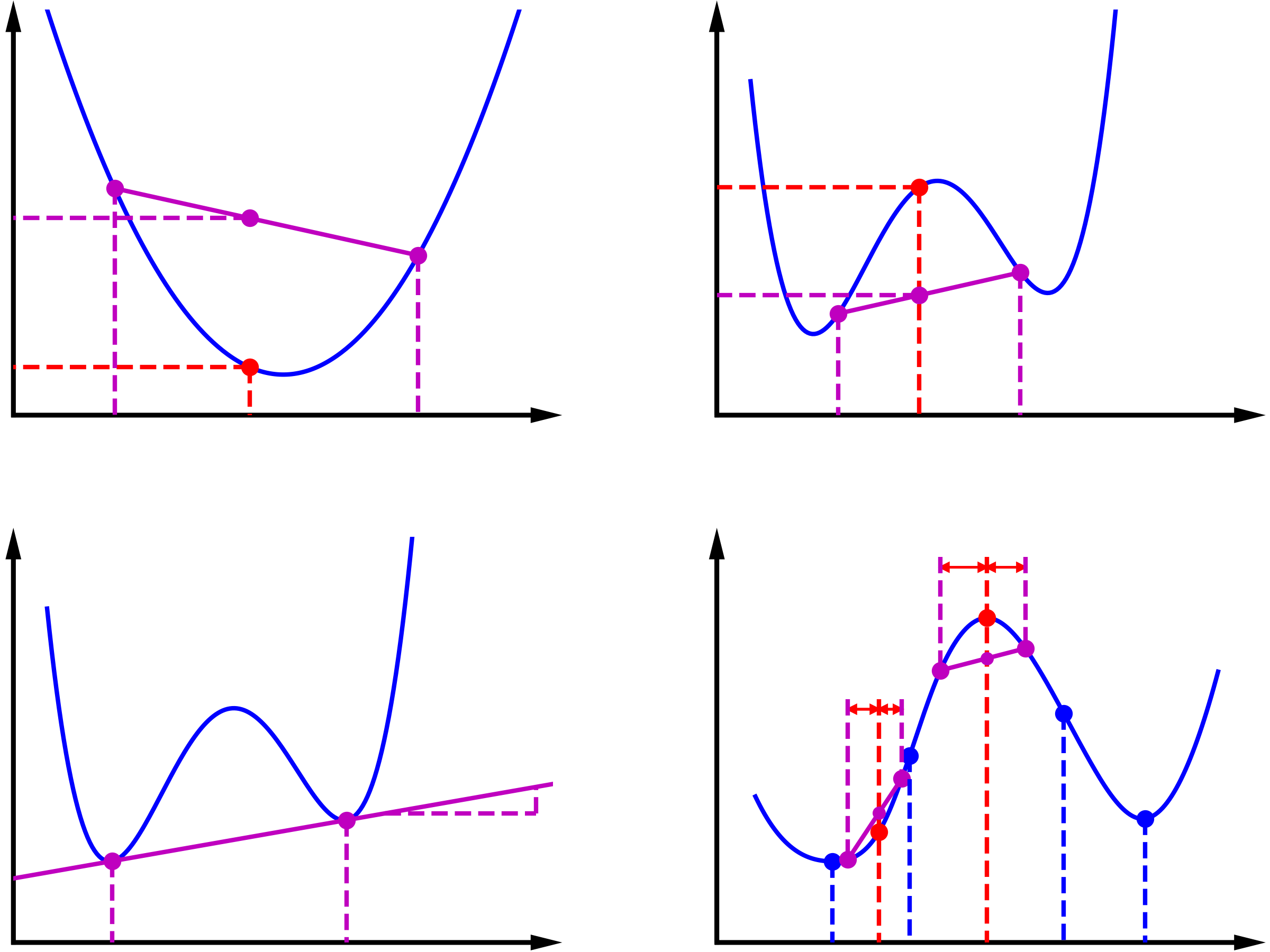}};
    \draw (-1.5,5) node[anchor=south west] {\bf (a)};
    
    \draw (-6.25,4.75) node[above] {$f(\rho)$};
    \draw (-6.25,-0.55) node[above] {$f(\rho)$};
    \draw (0.85,4.75) node[above] {$f(\rho)$};
    \draw (0.85,-0.55) node[above] {$f(\rho)$};
    
    \draw (-0.8,0.6) node[right] {$\rho$};
    \draw (-0.8,-4.7) node[right] {$\rho$};
    \draw (6.3,0.6) node[right] {$\rho$};
    \draw (6.3,-4.7) node[right] {$\rho$};
    
     \draw (-6,2.6) node[left,yshift=.2cm] {$\magenta{\begin{array}{c}\alpha f(\rho_1) + \\ (1-\alpha)f(\rho_2) \\ \end{array}}$};
    %\draw (-6.2,2.6) node[left] { $\magenta{\alpha f(\rho_1)+(1-\alpha)f(\rho_2)}$};
    \draw (-6.2,1.1) node[left] {$\red{f(\rho_0)}$};
    
    \draw (-5.2,0.6) node[below] {$\magenta{\rho_1}$};
    \draw (-3.8,0.6) node[below] {$\red{\rho_0}$};
    \draw (-2.1,0.6) node[below] {$\magenta{\rho_2}$};
    
    \draw (5.5,5) node[anchor=south west] {\bf (b)};
    
    \draw (2.1,0.6) node[below] {$\magenta{\rho_1}$};
    \draw (2.9,0.6) node[below] {$\red{\rho_0}$};
    \draw (3.9,0.6) node[below] {$\magenta{\rho_2}$};
    
    \draw (-1.5,-0.5) node[anchor=south west] {\bf (c)};
    
    \draw (-5.25,-4.7) node[below] {$\magenta{\rho_g}$};
    \draw (-2.8,-4.7) node[below] {$\magenta{\rho_\ell}$};
    \draw (-6.2,-4.1) node[left] {$\magenta{-p_T}$};
    \draw (-1.7,-3.3) node[below] {$\magenta{1}$};
    \draw (-0.8,-3.2) node[right] {$\magenta{\mu_T}$};
    
    \draw (5.5,-0.5) node[anchor=south west] {\bf (d)};
    
    \draw (2.0,-4.7) node[below] {$\blue{\rho_g}$};
    \draw (2.5,-4.7) node[below] {$\red{\rho_0^n}$};
    \draw (2.95,-4.7) node[below] {$\blue{\rho_g^s}$};
    \draw (3.5,-4.7) node[below] {$\red{\rho_0^s}$};
    \draw (4.4,-4.7) node[below] {$\blue{\rho_\ell^s}$};
    \draw (5.2,-4.7) node[below] {$\blue{\rho_\ell}$};
    
    \draw (2,-2.3) node[above] {$\red{\delta \rho_1}$};
    \draw (2.7,-2.3) node[above] {$\red{\delta \rho_2}$};
    
    \draw (3.2,-0.9) node[above] {$\red{\delta \rho_1}$};
    \draw (3.9,-0.9) node[above] {$\red{\delta \rho_2}$};
    \end{tikzpicture}}\fi
    \includegraphics[width=\textwidth]{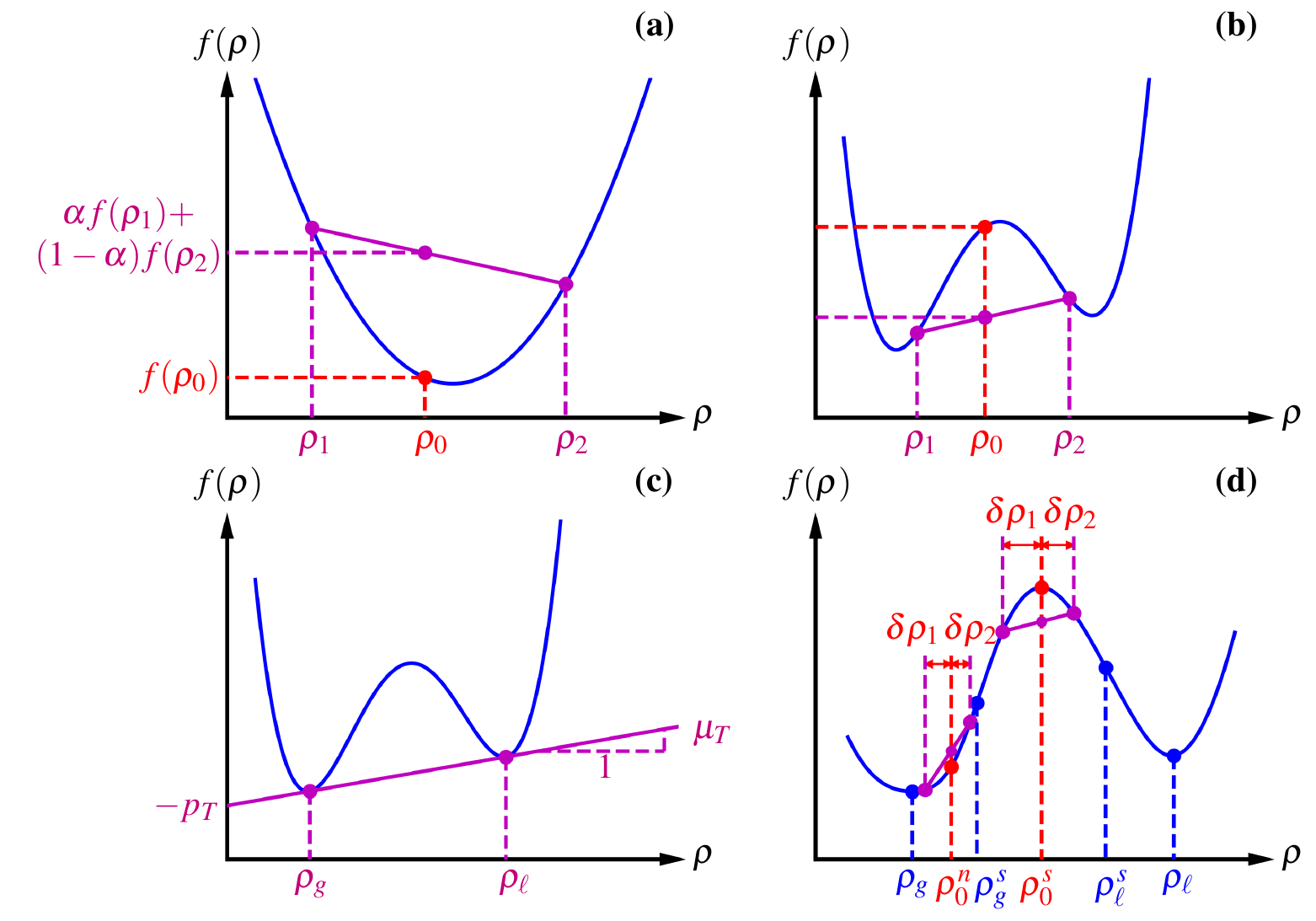}

    \caption{Analysis of phase separation based on the free-energy density $f(\rho)$. {\bf (a)} When the free energy density is convex, $f(\rho_0)$ is always lower than the free-energy density $\alpha f(\rho_1)+(1-\alpha)f(\rho_2)$ resulting from phase-separating between two densities $\rho_1$ and $\rho_2$ such that $\alpha\rho_1+(1-\alpha)\rho_2=\rho_0$. {(\bf b)} In a concave region of $f(\rho)$, the converse holds and phase separation is favored. {\bf (c)} The most favorable phase-separated state involves $\rho_g$ and $\rho_\ell$, which are the intersects between $f(\rho)$ and the linear section of its convex envelope, and delimit the region where the two do not overlap. The magenta line is called the common tangent to $f(\rho)$, its slope corresponds to the thermodynamic chemical potential $\mu_T$ while its intersect with the ordinate axis is given by the opposite of the thermodynamic pressure, $p_T$. {\bf (d)} By comparing the free energy density of a uniform profile at density $\rho_0$ and that of a coexistence state between densities $\rho_0-\delta \rho_1$ and $\rho_0+\delta \rho_2$, one predicts that phase-separation is triggered by a linear instability whenever $f''(\rho_0)<0$, whereas nucleation is required in $[\rho_g,\rho_g^s]\cup [\rho_\ell^s,\rho_\ell]$.}
    \label{fig:CTC}
\end{figure}

Let us now use the expressions~\eqref{eq:FEloc} and~\eqref{eq:FEDloc} to characterize the possibility of phase separation. When $\mu=\frac{\delta \mathcal{F}}{\delta \rho}$, the functional Fokker-Planck equation associated to~\eqref{eq:coldyn} admits a flux-free steady-state solution $P[\rho]=Z^{-1}\exp(-\mathcal{F}[\rho])$. The most likely profiles are thus those which minimize ${\cal F}[\rho]$. For a fixed averaged density $\rho_0$, one should compare the free energy of the homogeneous profile to that of a phase separated profile between a gas at some density $\rho_1$ and a liquid at density $\rho_2$, each occupying a fraction $\alpha$ and $1-\alpha$ of the system, with mass conservation imposing that $\rho_0=\alpha \rho_1 + (1-\alpha) \rho_2$. The uniform profile then has free energy  $\mathcal{V} f(\rho_0)={\mathcal{V}} f(\alpha \rho_1 + (1-\alpha) \rho_2)$, with $\mathcal{V}$ the volume of the system, while the phase-separated profile has free energy $\mathcal{V}[\alpha f(\rho_1)+(1-\alpha) f(\rho_2)]$.
Phase separation is thus statistically favored whenever
\begin{equation}\label{eq:PS}
    \alpha f(\rho_1)+(1-\alpha) f(\rho_2) < f(\alpha \rho_1 + (1-\alpha) \rho_2)\;, 
\end{equation}
which requires the function $f(\rho)$ to be non-convex. On the contrary, when $f(\rho)$ is convex, the uniform profile is always statistically favored. The two situations are illustrated in Fig.~\ref{fig:CTC}a-b. 

Equation~\eqref{eq:PS} may be satisfied by many different pairs
$(\rho_1,\rho_2)$ and a natural question is then which of these will
be statistically favoured. The answer comes from constructing the
convex envelope $f_c(\rho)$ of $f(\rho)$.  Regions in which
$f(\rho_0)=f_c(\rho_0)$ correspond to uniform phases. On the contrary,
the system is phase-separated when $f(\rho)>f_c(\rho)$. In such a
region, $f_c(\rho)$ is a linear function interpolating between two
densities such that $f(\rho_g)=f_c(\rho_g)$ and
$f(\rho_\ell)=f_c(\rho_\ell)$. The favoured solution within the region
is then a phase-separated state between $\rho_g$ and $\rho_\ell$,
where the relative fraction of each phase is set by the lever rule
$\rho_0=\alpha \rho_g + (1-\alpha) \rho_\ell$.  Geometrically, the
liquid and gas densities are found by using the so-called
common-tangent construction depicted in Fig.~\ref{fig:CTC}c.

Within the local approximation, we have thus established a direct
mapping between our active system and a passive equilibrium, which
allows discussing the existence of thermodynamic state
variables. Indeed, the common-tangent construction, derived above from
statistical arguments, also offers a thermodynamic interpretation. The
fact that the slopes of $f(\rho)$ are the same at $\rho_g$ and
$\rho_\ell$ can be read as an equality of thermodynamic chemical
potential, defined as $\mu_T=\frac{\partial {\cal F}}{\partial N}$,
where $N$ is the total number of particles. Using that $\mathcal
{F}\simeq \mathcal{V} f(\rho_0)$ and $\rho_0=N/\mathcal{V}$ indeed
leads to $\mu_T=\mathcal{V} f'(N/\mathcal{V})/\mathcal{V}=f'(\rho_0)$,
which allows identifying the value of field $\mu(r,[\rho])$ in
homogeneous phases with its thermodynamic counterpart in the
macroscopic limit. Similarly, the thermodynamic pressure is defined as
$p_T=-\frac{\partial \mathcal{F}}{\partial
  \mathcal{V}}=-f(\rho_0)+\rho_0 f'(\rho_0) $. The fact that the
tangents to $f(\rho)$ at $\rho_g$ and $\rho_\ell$ have the same
intersect with the ordinate axis thus conveys the equality of
thermodynamic pressure between the two phases.

The above discussion on the convexity of $f(\rho_0)$ provides an
intuitive understanding of the origin of phase separation and of the
selection mechanism for the coexisting densities. For the sake of
completeness, we mention that the mathematical derivation of the
common-tangent construction is more easily obtained by minimizing the
free energy constrained by the average density. To do so, consider the
function
\begin{equation}
  G(\alpha,\mu,\rho_g,\rho_\ell)=\alpha f(\rho_g)+(1-\alpha)f(\rho_\ell)-\mu[\alpha \rho_g + (1-\alpha) \rho_\ell-\rho_0]\;,
\end{equation}
where $\mu$ plays the role of a Lagrange multiplier. Minimizing $G$ with
respect to $\rho_g$ and $\rho_\ell$ leads to the equality of chemical
potentials, $\mu=f'(\rho_g)=f'(\rho_\ell)$. Then, minimizing $G$ with
respect to $\alpha$ leads to the equality of pressures
$f(\rho_g)-\rho_gf'(\rho_g)=f(\rho_\ell)-\rho_\ell f'(\rho_\ell)$. Minimizing
$G$ with respect to $\mu$ finally imposes the lever rule
$\alpha=(\rho_0-\rho_\ell)/(\rho_g-\rho_\ell)$.

From $f(\rho)$ we can thus predict whether the steady state of the system
is phase separated. How then does the system dynamically reach the
phase-separated state starting from an homogeneous one? A simple
answer comes from looking at the role of small fluctuations: Would
phase-separating between $\rho_0-\delta\rho_1$ and
$\rho_0+\delta\rho_2$ lower the free energy? If so, then small
fluctuations immediately trigger an instability ultimately leading to
phase separation, which is indeed the case whenever $f''(\rho_0)<0$
{\it i.e.} when the free energy is locally concave. This phenomenon is
termed spinodal decomposition and happens between the gas and liquid
spinodals that we denote $\rho^s_g$ and $\rho^s_\ell$. Using the
explicit expression of $f(\rho)$ given in Eq.~\eqref{eq:FEDloc}, the
spinodal region satisfies
 \begin{equation}\label{eq:spinodal}
     \frac{1}{\rho_0} + \frac{v'(\rho_0)}{v(\rho_0)}<0\;.
 \end{equation}
Interestingly, this is exactly the criterion~\eqref{eq:criterion} derived heuristically in Section~\ref{sec:handwaving}. On the contrary in $[\rho_g,\rho_g^s]\cup[\rho_\ell^s,\rho_\ell]$ the system is globally unstable but linear fluctuations will not suffice to lead to phase separation. In these region, nucleation is required to destabilize an initially homogeneous profile.

The local approximation is appealing because it offers a simple
framework to predict the occurrence of motility-induced phase
separation and captures quantitatively the linear instability
criterion~\eqref{eq:spinodal}. However, it suffers from a major
drawback: the measure corresponding to the free
energy~\eqref{eq:FEloc} is factorized and thus predicts all points in
space to be statistically independent. There is thus no way to
understand the physics of domain walls or coarsening within this
framework, nor to account for any spatial structure exhibited by the
system. To fix this, a natural strategy is to go beyond the local
approximation $\tilde\rho(\bfr)=\rho(\bfr)$. This is the purpose of
the next subsection which includes higher-order gradient terms. As we
shall see, the mapping to equilibrium then breaks down and the
connection to thermodynamics is lost. While the equality of `chemical
potentials' between coexisting phases is preserved, this is not true
of $p_T$: because of gradient terms, the common-tangent construction will be replaced
by an uncommon-tangent construction~\cite{wittkowski2014scalar}.

\subsection{Gradient expansion: beyond the local approximation}
The next order expansion of $v(r,[\rho])$ is obtained by using Eq.~\eqref{eq:tilderho} to write $\tilde\rho=\rho+\frac{D}{\gamma} \Delta \rho$, which leads to a `quasi-local' approximation
\begin{equation}\label{eq:quasiloc}
    v(r,[\rho])\simeq v(\rho(\bfr))+v'(\rho(\bfr))\frac{D}{\gamma} \Delta \rho(\bfr)\;.
\end{equation}
In turn, this leads to the generalized chemical potential
\begin{equation}\label{eq:muNL}
    \mu(\bfr,[\rho])=\ln[\rho(\bfr)v(\rho(\bfr))]-\kappa(\rho(\bfr))\Delta\rho\quad\text{with}\quad\kappa(\rho(\bfr))=-\frac{D v'(\rho(\bfr))}{\gamma v(\rho(\bfr))}\;.
\end{equation}

\subsubsection{Equilibrium mapping}
The local part of $\mu$ in Eq.~\eqref{eq:muNL} is nothing but the local theory developed in the previous subsection. It can be integrated functionally to lead to the free energy density~\eqref{eq:FEDloc}. 

To check the conditions under which $-\kappa(\rho) \Delta \rho$ is a functional derivative, we use Eq.~\eqref{eq:D} to compute its contribution to $\mathcal{D}(\bfr,\bfr')$, which leads to
\begin{equation}
    \mathcal{D}(\bfr,\bfr')=\kappa(\rho(\bfr'))\Delta_\bfr \delta(\bfr-\bfr')-\kappa(\rho(\bfr))\Delta_{\bfr'} \delta(\bfr-\bfr')\;.
\end{equation}
Integrating against pairs of function $\phi(\bfr)$ and $\psi(\bfr')$ then leads to
\begin{equation}
    \int d\bfr d\bfr' \mathcal{D}(\bfr,\bfr')\phi(\bfr)\psi(\bfr')=\int d\bfr \nabla \kappa(\rho(\bfr)) \cdot [\phi(\bfr) \grad \psi(\bfr)-\psi(\bfr)\grad\phi(\bfr)]\;.
\end{equation}
This vanishes for any $\rho, \psi, \phi$ if and only if $\kappa'(\rho)=0$. Using the expression~\eqref{eq:muNL} for $\kappa$, this shows that the quasi-local approximation~\eqref{eq:quasiloc} admits an equilibrium free energy iff 
\begin{equation}\label{eq:vofrhoexp}
    v(\tilde\rho)=v_0 \exp[\lambda \tilde\rho]\;.
\end{equation}
This functional form of $v$ leads to MIPS for $\lambda<0$ and to a free energy
\begin{equation}\label{eq:FEexpgrad}
    {\cal F}=\int d\bfr \Big[ \rho(\bfr)\ln\rho(\bfr) +\frac{\lambda}{2} \rho(\bfr)^2-\frac{\lambda D}{2\gamma} [\nabla \rho(\bfr)]^2\Big ]\;.
\end{equation}
While interfaces now have a finite cost, the free energy density in Eq.~\eqref{eq:FEexpgrad} is not bounded from below. As $\rho(\bfr)$ keeps increasing locally, $f(\rho(\bfr))$ becomes arbitrarily negative: a macroscopic fraction of the mass thus tends to accumulate in a finite portion of space. For the functional form~\eqref{eq:vofrhoexp}, our equilibrium theory thus predicts --- to the order~\eqref{eq:quasiloc} in the gradient expansion --- a condensation transition instead of a phase separation between finite densities.

For more general forms of $v(\rho)$, the system is endowed with a \textit{bona fide} nonequilibrium dynamics. A local approximation to the chemical potential may qualitatively explain the occurrence of MIPS, but more general properties, such as the existence of state functions or simply coexisting densities appear out of reach.

\subsubsection{A generalized equilibrium theory at mean-field level}
Progress can however be made at mean-field level. Introducing $\bar \rho=\langle \rho\rangle$ and making the local mean-field approximation $\langle g(\bar \rho) \rangle=g(\langle\bar \rho \rangle)$ for any function $g$, the dynamics of $\bar \rho$ can be obtained as:
\begin{equation}
    \partial_t \bar\rho(\bfr,t)=\nabla \cdot [ M \nabla \bar \mu]\qquad\text{with}\qquad\bar\mu=\log[\bar\rho v(\bar\rho)]-\kappa(\bar\rho)\Delta\bar\rho\;.
\end{equation}
While $\bar\mu$ cannot be written as the functional derivative of a functional ${\cal F}$ with respect to $\rho$, we can look for a change of variable $R(\bar\rho)$ such that 
\begin{equation}\label{eq:muH}
    \bar\mu = \frac{\delta \mathcal{H}}{\delta R(\bar\rho)} \;.
\end{equation}
Remarkably, defining $R$ from
\begin{equation}
    R'(\bar\rho)=-\frac{1}{\kappa(\bar\rho)}
\end{equation}
indeed leads to Eq.~\eqref{eq:muH}, where the generalized free energy $\mathcal{H}$ is given by~\cite{solon2018PRE,solon2018njp}
\begin{equation}\label{eq:H}
    \mathcal{H}=\int d\bfr \left[\phi(R)+\frac{\kappa}{2 R'(\bar\rho)} (\nabla R)^2\right] \quad\text{where}\quad\frac{d \phi(R)}{dR}=\ln[\bar\rho v(\bar\rho)]\;.
\end{equation}
Note that we here provide directly the change of variable
$R(\bar\rho)$ that allows writing Eq.~\eqref{eq:muH}. This change of
variable can actually be constructed systematically by rewriting
Eq.~\eqref{eq:D} using functional derivatives with respect to
$R(\bfr')$ and $R(\bfr)$, and by requiring the corresponding distribution
$D$ to vanish~\cite{JeremInPrep}.

\begin{figure}
    \centering
    \includegraphics[width=6cm]{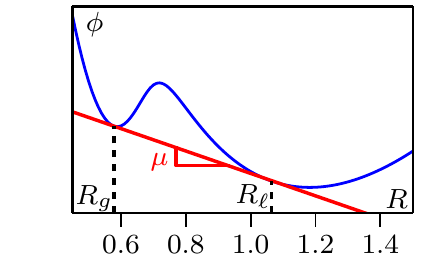}\hspace{1cm}
        \includegraphics{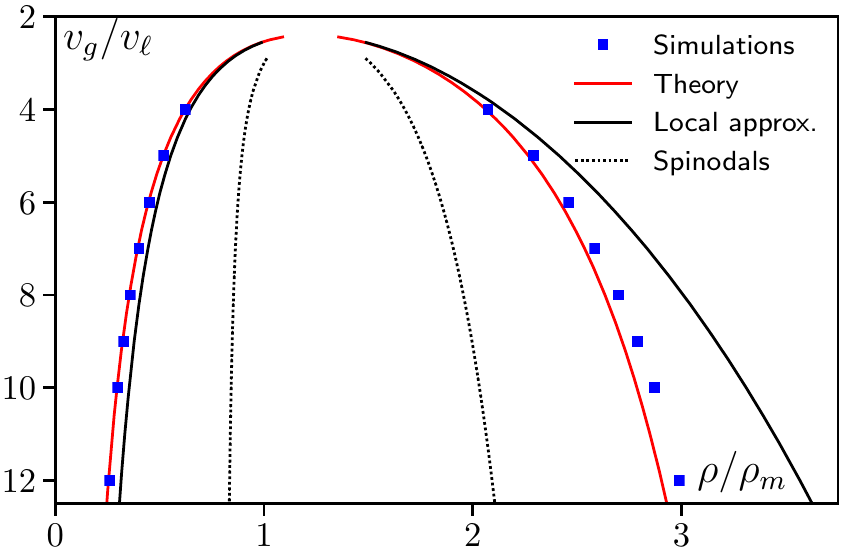}
    \caption{(a) The binodals are computed through a common tangent construction on the generalized free energy defined in Eq.~\eqref{eq:H}. (b) Comparison between the binodals measured in simulations of RTPs in 2d and the binodals predicted by the local approximation (plain black line) and the mean-field non-local theory (red line). Note that we do not expect the latter to be a good approximation close to the critical point where fluctuations are crucial and at high $v_0$ because gradients become steeper and higher-order gradient terms become important. Parameters: the self-propulsion given by Eq.~\eqref{eq:vofrhoNJP} with $\rho_m=200$, $v_\ell=5$, $\tau=1$ and system size $50\times 50$.}
    \label{fig:generalizedthermo}
\end{figure}

Once we have derived the effective free energy~\eqref{eq:H}, constructing the steady-state profiles $R(\bfr)$ amounts to an equilibrium problem. For instance, the coexisting generalized densities $R_g$ and $R_\ell$ can be
obtained from a common tangent construction on $\phi(R)$. Inverting
$R(\rho)$ then leads to the values of the gas and liquid binodals
$\rho_g$ and $\rho_\ell$. This procedure is illustrated in
Fig.~\ref{fig:generalizedthermo} where we compare with results from
microscopic simulations using a self-propulsion velocity given by
\begin{equation}\label{eq:vofrhoNJP}
    v(\tilde\rho)=v_g+\frac{v_    \ell-v_g}2 \left[1+\tanh\left(2\frac{\tilde\rho}{\rho_m}-2\right)\right]
\end{equation}
and a coarse-graining kernel
\begin{equation}
    G(\bfr)=\frac{\Theta(r_0-|\bfr|)}{Z} \exp\left(-\frac{r_0^2}{r_0^2-r^2}\right)\;,
\end{equation}
where $\Theta$ is the Heaviside function.

The binodals predicted by the local, equilibrium theory correspond to the solid black line in Fig.~\ref{fig:generalizedthermo}b and clearly fail to describe the microscopic simulations given by the symbols. On the contrary, the procedure described above, summarized in Fig.~\ref{fig:generalizedthermo}a, lead to the solid red lines, which agree very well with the microscopic simulations. 

\subsection{The fully non-local theory}\label{sec:nonlocal}
Finally, a natural question is whether a mapping to equilibrium exists for the general form~\eqref{eq:vconvol}, where $v$ is a functional of the non-local effective density $\tilde \rho(\bfr)=\int d\bfr' G(\bfr-\bfr') \rho(\bfr')$. In such a case, the distribution $\mathcal{D}$ is equal to
\begin{equation}
    \mathcal{D}(\bfr,\bfr')=\frac{1}{v(\bfr,[\rho])}\frac{\delta v(\bfr,[\rho])}{\delta \rho(\bfr')}-\frac{1}{v(\bfr',[\rho])}\frac{\delta v(\bfr',[\rho])}{\delta \rho(\bfr)}\;.
\end{equation}
Using that Eq.~\eqref{eq:vconvol} can be written as $v(\bfr,[\rho])=v[\int d\bfr'' G(\bfr-\bfr'') \rho(\bfr'')]$, one gets
\begin{equation}
    \frac{\delta v(\bfr,[\rho])}{\delta \rho(\bfr')}= v'(\bfr,[\rho])\int d\bfr'' G(\bfr-\bfr'') \frac{\delta \rho(\bfr'')}{\delta \rho(\bfr')}\;,
\end{equation}
where $v'(\bfr,[\rho])=\frac{d}{d\tilde \rho} v(\tilde \rho)$. In turns, this leads to
\begin{equation}\label{eq:Dgene}
    \mathcal{D}(\bfr,\bfr')=\frac{v'(\bfr,[\rho])}{v(\bfr,[\rho])}G(\bfr-\bfr')-\frac{v'(\bfr',[\rho])}{v(\bfr',[\rho])}G(\bfr'-\bfr)\;.
\end{equation}
Given the isotropy of $G$, the solution to Eq.~\eqref{eq:Dgene} is
\begin{equation}
    \frac{v'(\bfr',[\rho])}{v(\bfr',[\rho])}=\lambda \in \mathbb{R}\qquad\text{so that}\qquad v(\bfr,[\rho])=v_0 e^{\lambda \int d\bfr' G(\bfr-\bfr') \rho(\bfr')}\;.
\end{equation}
The sole function $v(\tilde\rho)$ leading to an equilibrium mapping is thus the exponential function~\cite{tailleur_statistical_2008,grafke2017spatiotemporal}. In this case, the free energy ${\cal F}$ is given by
\begin{equation}\label{eq:mappingfull}
    \mathcal{F}=\int d\bfr \rho(\bfr)\ln[\rho(\bfr)] + \frac{1}{2} \lambda\int d\bfr d\bfr' \rho(\bfr)\rho(\bfr') G(\bfr-\bfr')\;.
\end{equation}
The dynamics of $N$ RTPs interacting via quorum-sensing interactions and a density-dependent swim speed that varies exponentially thus map, at the coarse-grained level, to that of $N$ equilibrium Brownian colloids whose interaction potential is given by $\lambda G(\bfr)$. If $\lambda G(\bfr)$ corresponds, say, to a repulsive hard core and attractive tails, then MIPS in QSAPs is mapped exactly to the liquid-gas phase separation observed for attractive colloidal particles.

The mapping discussed in this section belongs to a larger class of mappings between active and passive particles~\cite{o2020lamellar} that also exist for tactic dynamics. The mapping~\eqref{eq:mappingfull} does not simply map MIPS onto the phase-separation observed for passive particles, it says that the full range of phenomena observed for interacting passive particles can be observed for active ones, upon a highly counter-intuitive mapping between the kernel $G$ and the interaction potential of passive particles.  

\subsection{QSAPs: conclusion}
We close here our discussion of active particles interacting via quorum-sensing interactions, which is the model for which MIPS is best understood. Depending on the question asked, various levels of description offer answers of increasing complexities. First, a simple heuristic argument, detailed in Section~\ref{sec:handwaving}, explains the instability mechanism leading to MIPS. Then, a qualitative mapping to equilibrium models, discussed in Section~\ref{sec:local}, establishes an unexpected similarity between motility regulation in active matter and pairwise forces in equilibrium systems. This mapping is shown to break down at the first non-trivial order in a gradient expansion. Then a generalized free energy can be constructed at mean-field level, whose predictive power was demonstrated numerically. Note that, in addition to the coexisting densities discussed so far, this generalized free energy allows tackling a broader set of questions. For instance, it allows demonstrating that the thermodynamic pressure $p_T$ is not a state function and instead suggests $p_R=R \frac{d \phi}{d R}-\phi$ as a (mean-field) alternate. The latter can be shown, for instance, to govern Laplace pressure jumps that account for finite-size corrections to coexisting binodals in finite systems~\cite{solon2018njp}.

More generally, the exact mapping between active and passive systems established at the fluctuating-hydrodynamics level in Section~\ref{sec:nonlocal} paves the way to a thermodynamics of QSAPs. It comes with restrictions and cannot encompass all scalar active matter systems, but it offers a route to thermodynamics where everything is under control analytically.

In the next section, we briefly discuss the case of active particles interacting via pairwise forces, highlighting similarities and differences with the case of QSAPs.

\section{Active particles interacting via pairwise forces}
\label{sec:PFAPs}

As discussed in the introduction, another important class of systems that undergo motility-induced phase separation consists in self-propelled particles interacting via repulsive forces (PFAPs). Their dynamics can typically be written as
\begin{equation}\label{eq:PFAPs}
    \dot \bfr_i=v_0 \bfu_i - \mathfrak{u} \sum_{j\neq i} \nabla_i V(\bfr_i-\bfr_j)\;, 
\end{equation}
where $v_0$ is the single-particle propulsion speed, $\bfu_i$ the orientation of particle $i$, $\mathfrak{u}$ the particle mobility, and $V$ an interaction potential. 
MIPS has been reported for a variety of repulsive potentials such as harmonic spheres~\cite{Fily2012PRL}, the Yukawa potential~\cite{buttinoni_dynamical_2013} or the Week-Chandler-Andersen potential~\cite{redner2013structure}. While most studies have focused on the 2D case, MIPS has also been reported in three dimensions~\cite{stenhammar2014phase,wysocki2014cooperative,turci2021phase} and studied analytically in the limit of infinite space dimensions~\cite{de2019active}. 

\subsection{A quorum-sensing approximation}
Let us now illustrate how the emergence of MIPS in PFAPs can easily be understood using a local mean-field theory. Consider the dynamics~\eqref{eq:PFAPs} for particle $i$. Introducing the projector on the orientation $\bfu_i$, whose components are given by $u_i^\alpha u_i^\beta$, the dynamics can be rewritten as
\begin{equation}\label{eq:PFAPseff}
    \dot \bfr_i=v_{i,\rm eff} \,\bfu_i +\boldsymbol\xi_i\;,
\end{equation}
where we have introduced the effective speed $v_{i,\rm eff}$ and the `noise'  $\boldsymbol\xi_i$, defined by
\begin{equation}\label{eq:PFAPscoef}
\boldsymbol\xi_i\equiv - \mathfrak{u}\sum_{j\neq i} (1-\bfu_i \bfu_i) \nabla_i V(\bfr_i-\bfr_j)\qquad\text{and}\qquad v_{i,\rm eff}\equiv v_0-\mathfrak{u}\sum_{j\neq i} \bfu_i \cdot \nabla V(\bfr_i-\bfr_j)\;.
\end{equation}
Equations~\eqref{eq:PFAPseff} and~\eqref{eq:PFAPscoef} are exact; they simply amount to splitting the forces exerted by the particles $j\neq i$ on particle $i$ between a contribution parallel to the orientation ${\bfu_i}$, which renormalizes $v_0$ into $v_{i,\rm eff}$, and a contribution orthogonal to $\bfu_i$ which we denote $\boldsymbol{\xi}_i$. 

Let us now employ a local mean-field approximation to get more insight into the behaviors of $v_{i,\rm eff}$ and $\boldsymbol{\xi}_i$. In a homogeneous phase, the orientation ${\bfu_i}$ is an axis of symmetry of particle $i$ so that $\langle \boldsymbol\xi_i\rangle_j=0$, where $\langle\dots\rangle_j$ refers to an average over $\bfr_j$, keeping $\bfr_i$ and $\bfu_i$ fixed. On the contrary, most collisions between particle $i$ and other particles happen when the latter are ahead of particle $i$~\cite{bialke2013microscopic}. This can be easily understood by considering the situation in which you are running, say, to catch a train: most of the time, you will collide with people ahead of you, not behind you, especially when everyone is running in random directions. The result is that $\delta v_i\equiv -\mathfrak{u} \langle \sum_{j\neq i} \bfu_i \cdot \nabla V(\bfr_i-\bfr_j) \rangle_j$ is negative: repulsive forces oppose self-propulsion. In the low-density limit, where particles are weakly correlated, $\delta v$ is simply proportional to the number of particles $j$ with which particle $i$ interact, and thus to the local density. Furthermore, $-\nabla V$ is of the order of the typical force to stall a particle, and thus $|\mathfrak{u} \nabla V|\simeq v_0$. Overall, $\delta v_i \propto \rho(\bfr_i) v_0$. Introducing a proportionality constant $1/\rho^*$, this leads to an average propulsion speed that scales as
\begin{equation}\label{eq:PFAPSeffspeed}
    v_{\rm eff}(\bfr_i)\equiv \langle v_{i,\rm eff}\rangle=v_0[1-\rho(\bfr_i)/\rho^*]\;,
\end{equation}
where $\rho^*$ is \textit{a priori} unknown and has the dimension of a density. This linear scaling has indeed been measured in simulations~\cite{Fily2012PRL} and derived analytically in infinite dimensions~\cite{de2019active}. 

Given the above discussion, it is tempting to approximate PFAPs by QSAPs whose self-propulsion speeds would be given by Eq.~\eqref{eq:PFAPSeffspeed}~\cite{Fily2012PRL,bialke2013microscopic,stenhammar2013continuum}. An appealing feature of this approximation is that it offers a simple mechanism accounting for MIPS: for QSAPs, the  criterion~\eqref{eq:criterion} applied to the self-propulsion speed~\eqref{eq:PFAPSeffspeed} predicts  a linear instability leading to MIPS whenever 
\begin{equation}\label{eq:critlinear}
    \rho_0>\rho^*/2\;.
\end{equation}
This quorum-sensing approximation however suffers from important limitations. First, it predicts that the chemical potential $\mu=\ln[\rho v_{\rm eff}(\rho)]$ should be equal in coexisting phases, which is not verified numerically, as we discuss below. Then, the instability criterion~\eqref{eq:critlinear} is independent from $v_0$. It thus predicts a phase separation for arbitrarily small self-propulsion speeds $v_0$. This, again, contradicts numerical results, which show that the run-length $v_0/D_r$ has to be larger than $\sim 15$ particle size $r_0$~(see Fig.~\ref{fig:MIPSLJ}). Finally, the QSAPs approximation predicts a Laplace-pressure jump in finite-sized droplets coexisting with a gaseous phase that has an opposite sign with respect to the one measured for PFAPs~\cite{solon2018njp}. For all these reasons, an alternate route is preferable to study the MIPS observed for PFAPs. 

\subsection{The stress tensor of PFAPs}
We now briefly present the path laid out in~\cite{solon2015pressure,solon2018njp,martin_statistical_2021} for the case of ABPs. Let us denote  $\hat \rho(\bfr)=\sum_{i=1}^N \delta(\bfr-\bfr_i)$ the instantaneous, stochastic density field and $\rho(\bfr)=\langle \hat\rho(\bfr)\rangle$ its average over realizations of the active dynamics of ${\bfu_i(t)}$. Using the chain rule, one gets for the dynamics of $\rho(\bfr,t)$
\begin{equation}
    \partial_t \rho(\bfr,t)=-\nabla \cdot \bfJ(\bfr,t)\;,
\end{equation}
where the average density current $\bfJ$ can be split between active and passive contributions: $\bfJ=\bfJ^a+\bfJ^p$, with
\begin{equation}\label{eq:currentsPFAPs}
\bfJ^a(\bfr,t)=\langle\sum_{i=1}^N v_0 \bfu_i\delta(\bfr-\bfr_i)\rangle\qquad\text{and}\qquad \bfJ^p(\bfr,t)= - \mathfrak{u} \int d\bfr' \langle \rho(\bfr)\rho(\bfr')\rangle \grad V(\bfr-\bfr')\;.
\end{equation}
These two currents account for the displacements of particles due to their self-propulsion and to the pairwise forces, respectively. 

Let us first note that $\bfJ^p$ would also contribute to the current for passive Brownian particles interacting via pairwise forces, although, in this case, the average in Eq.~\eqref{eq:currentsPFAPs} would be computed using the equilibrium Boltzmann weight. Despite this difference, we can directly borrow results derived for equilibrium systems by Irving and Kirkwood~\cite{irving1950statistical}, who showed that $\bfJ^p$ can be written as the divergence of a stress tensor: $\bfJ^p=\mathfrak{u} \nabla \cdot \boldsymbol{\sigma}^{IK}$ where
\begin{equation}\label{eq:sigmaIK}
{\sigma}^{IK}_{\alpha\beta}(\bfr)=\frac 1 2 \int d\bfr' \Big\{\frac{r'_\alpha r'_\beta}{|\bfr'|} \frac{d V(|\bfr'|)}{d|\bfr'|}\int_0^1 d\lambda \langle \hat \rho(\bfr+(1-\lambda)\bfr') \hat\rho(\bfr-\lambda\bfr')\rangle\Big\}\;. 
\end{equation}
While Eq.~\eqref{eq:sigmaIK} is not particularly illuminating, it can be shown~\cite{solon2015pressure} that $\boldsymbol{\sigma}^{IK}$ acquires a simpler meaning in homogeneous systems, where, for instance, $\sigma_{xx}^{IK}(x,y)$ measures the force density exerted by the particles at $(x'>x,y')$ on those at $(x'<x,y)$. 

Up to a factor of $\mathfrak{u}$, $\bfJ^a(\bfr)$ measures  the density of active forces experienced by the system at position $\bfr$. Using It\=o formula, its dynamics can be computed as:
\begin{equation}\label{eq:beautiful}
    \partial_t \bfJ^a= \langle \sum_{i=1}^N v_0 \dot \bfu_i  \delta(\bfr-\bfr_i)\rangle -\nabla_\bfr \cdot \langle \sum_{i=1}^N \dot \bfr_i v_0 \bfu_i  \delta(\bfr-\bfr_i)\rangle  - D_r \langle \sum_{i=1}^N v_0 \bfu_i  \delta(\bfr-\bfr_i)\rangle\;,
\end{equation}
In the steady-state, $\partial_t \bfJ^a$ and  $\langle \dot \bfu_i \rangle$ vanish. Using that the last term of Eq.~\eqref{eq:beautiful} is nothing but $-D_r \bfJ^a$, we find~\cite{fily2017mechanical}:
\begin{equation}\label{eq:Ja}
    {\bfJ^a}(\bfr)=\mathfrak{u} \nabla \cdot \sigma^a(\bfr)\;,\qquad\text{where}\qquad \sigma^a(\bfr)=-\langle \sum_i \dot \bfr_i \frac{v_0\bfu_i}{\mathfrak{u} D_r} \delta(\bfr-\bfr_i)\rangle\;.
\end{equation}
Inspection of Eq.~\eqref{eq:Ja} shows that the active stress tensor $\sigma^a$ measures the flux of the so-called `active impulse'~\cite{fily2017mechanical}: 
\begin{equation}\label{eq:activeimpulse}
    \delta {\bf p}_i^a=\frac{v_0 \bfu_i(t)}{\mathfrak{u} D_r}=\int_t^{\infty} ds \Big\langle \frac{v_0 \bfu_i(s)}{\mathfrak{u}} \Big\rangle\;,
\end{equation}
which is nothing but the average momentum an active particle will gain, between $t$ and the end of times,  due to its active force $v_0 \bfu_i/\mathfrak{u}$. 

Together, Eqs.~\eqref{eq:sigmaIK} and~\eqref{eq:Ja} show that the total current can be written as the divergence of a generalized stress tensor:
\begin{equation}\label{eq:JandSigma}
    \bfJ(\bfr)=\mathfrak{u} \nabla \cdot \sigma(\bfr)\qquad\text{where}\qquad \sigma(\bfr)=\sigma^{IK}(\bfr)+\sigma^a(\bfr)\;.
\end{equation} 
The mechanical implications of $\sigma$ become apparent by noting that, in the presence of an external potential $V_w(\bfr)$, the current becomes $\bfJ(\bfr)=\mathfrak{u} \nabla \cdot \sigma(\bfr) - \rho \nabla V_w(\bfr)$. When $V_w$ models a wall localized at $x=0$, which prevents the particles from entering the $x<0$ subspace, the mechanical pressure exerted by the active particles on this wall in a flux-free steady state can be measured as~\cite{solon2015nphys} 
\begin{equation}\label{eq:pressure}
    P_m=-\int_{-\infty}^{x_b} dx \rho(x,y_b) \partial_x V(x,y_b)=-\sigma_{xx}(x_b)\;,
\end{equation}
where $(x_b,y_b)$ corresponds to a point deep in the bulk of the system. Like passive particles, ABPs interacting via pairwise forces thus admit a stress tensor whose diagonal components are the opposite of the mechanical pressure exerted by the particles on confining walls.
    
\subsection{MIPS in PFAPs: A mechanical instability}
While the mechanical properties of active particles are a fascinating topic~\cite{takatori2014swim,yang2014aggregation,solon2015pressure,solon2015nphys,fily2017mechanical,dor2018forces}, we are here interested in the consequences of Eq.~\eqref{eq:JandSigma} for MIPS. Note that, so far, everything we have done is exact and we now employ approximations to relate MIPS to the properties of $\sigma$. In particular, we assume that $\sigma$ is well described, in homogeneous systems, by an equation of state $\sigma(\rho_0)$. (Remember that the density is the only hydrodynamic field in MIPS.)

Let us first look at the linear stability of a homogeneous profile at density $\rho_0$ against a fluctuation along the $x$ axis, $\delta\rho(\bfr,t)=\delta\rho_q(t) \cos(q x)$. The time evolution of $\delta\rho_q$ is then given by
\begin{equation}
\frac{d}{dt} \delta \rho_q(t)= q^2 {\mathfrak{u}}\sigma'_{xx}(\rho_0) \delta \rho_q(t)=-q^2 {\mathfrak{u}} P'_m(\rho_0) \delta \rho_q(t)\;,
\end{equation}
where we used Eq.~\eqref{eq:pressure} to work with the pressure instead of the stress tensor. Much like in a passive system, negative values for $P_m'(\rho_0)$ signal a mechanical instability. Let us now analyze which contributions to $P_m(\rho_0)$ can be responsible for such a decrease in mechanical pressure as $\rho_0$ increases.

The passive part of the stress tensor, which leads to a pressure
$P_m^{IK}(\rho_0)$, behaves qualitatively as that of a passive system
comprising repulsive particles: $P_m^{IK}(\rho_0)$ is an increasing
function that scales as $\rho_0^2$ at small densities due to the
pairwise nature of the forces and diverges at close
packing. Inspection of Eq.~\eqref{eq:sigmaIK} {suggests} the following
scaling form for $P_m^{IK}(\rho_0)$: it is proportional to the typical
distance between interacting particles multiplied by the interparticle
force, due to the factor
$\frac{r'_\alpha r'_\beta}{|\bfr'|} \frac{d
  V(|\bfr'|)}{d|\bfr'|}$, whereas the rest of the expression scales
as a density. As argued above, the typical scale of the
interparticle forces in MIPS is set by the self-propulsion they
oppose: $\frac{d V(|\bfr'|)}{d|\bfr'|}$ is, typically, a stalling
force and scales as $v_0/\mathfrak{u}$. The particle size $r_0$ then
sets the scale for the interparticle distance and we find
\begin{equation}\label{eq:PmIK}
    P_m^{IK}(\rho_0) \sim  \frac{v_0 r_0}{\mathfrak{u}} \rho_0\mathcal{S}\left(\frac{\rho_0}{\rho^*}\right)
\end{equation}
where $\mathcal{S}$ is a scaling function. At small density, the probability of encounters scale as $\rho_0^2$ so that $\mathcal{S}(u)\sim u$ as $u\to 0$. 

Using Eq.~\eqref{eq:PFAPSeffspeed}, we approximate $P_m^a(\rho_0)\equiv - \sigma^a_{xx}(\rho_0)$ as~\cite{solon2015pressure}
\begin{equation}\label{eq:Pma}
    P_m^a(\rho_0) \sim \frac{v_0 v_{\rm eff}}{\mathfrak{u} D_r} \rho_0=\frac{v_0^2}{\mathfrak{u} D_r} \rho_0 \left(1-\frac{\rho_0}{\rho^*}\right)\;. 
\end{equation}
While $P_m^a(\rho_0)$ is increasing---and hence stabilizing---at small densities, it decreases at larger ones due to the decrease in the effective speed. The active pressure can thus trigger an instability. Note that ${P_m^a}'(\rho_0)<0$ amounts to $\rho>\rho^*/2$: one recovers the "QSAPs" criterion~\eqref{eq:critlinear}. The latter accounts for the instability triggered by the slowing down but is missing the crucial contribution of $P_m^{IK}$.

\begin{figure}
  \centering
  \if{
    \begin{tikzpicture}
    \path (0,0) node {\includegraphics[height=4cm]{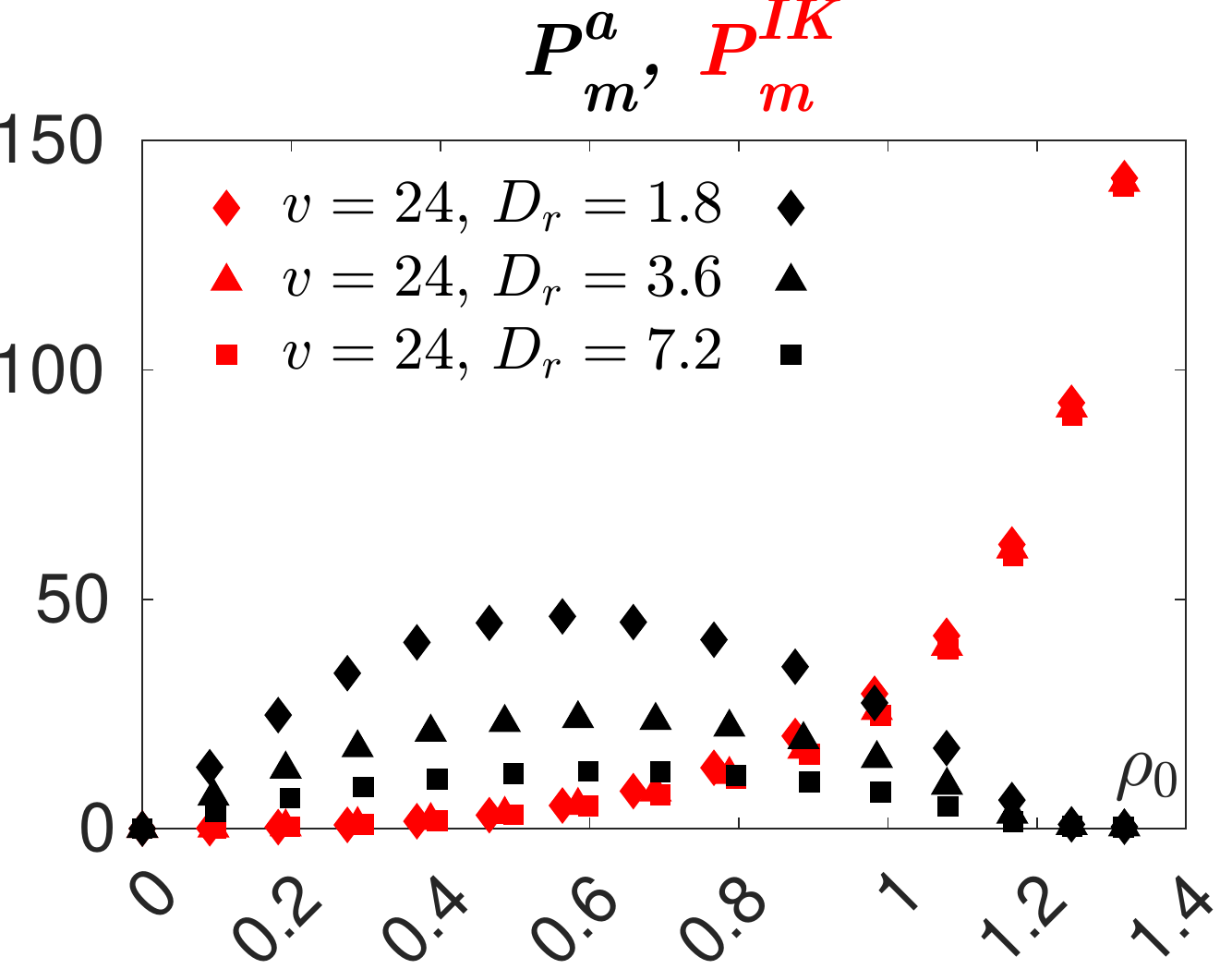}};
    \path (5.35cm,0) node {\includegraphics[height=4cm]{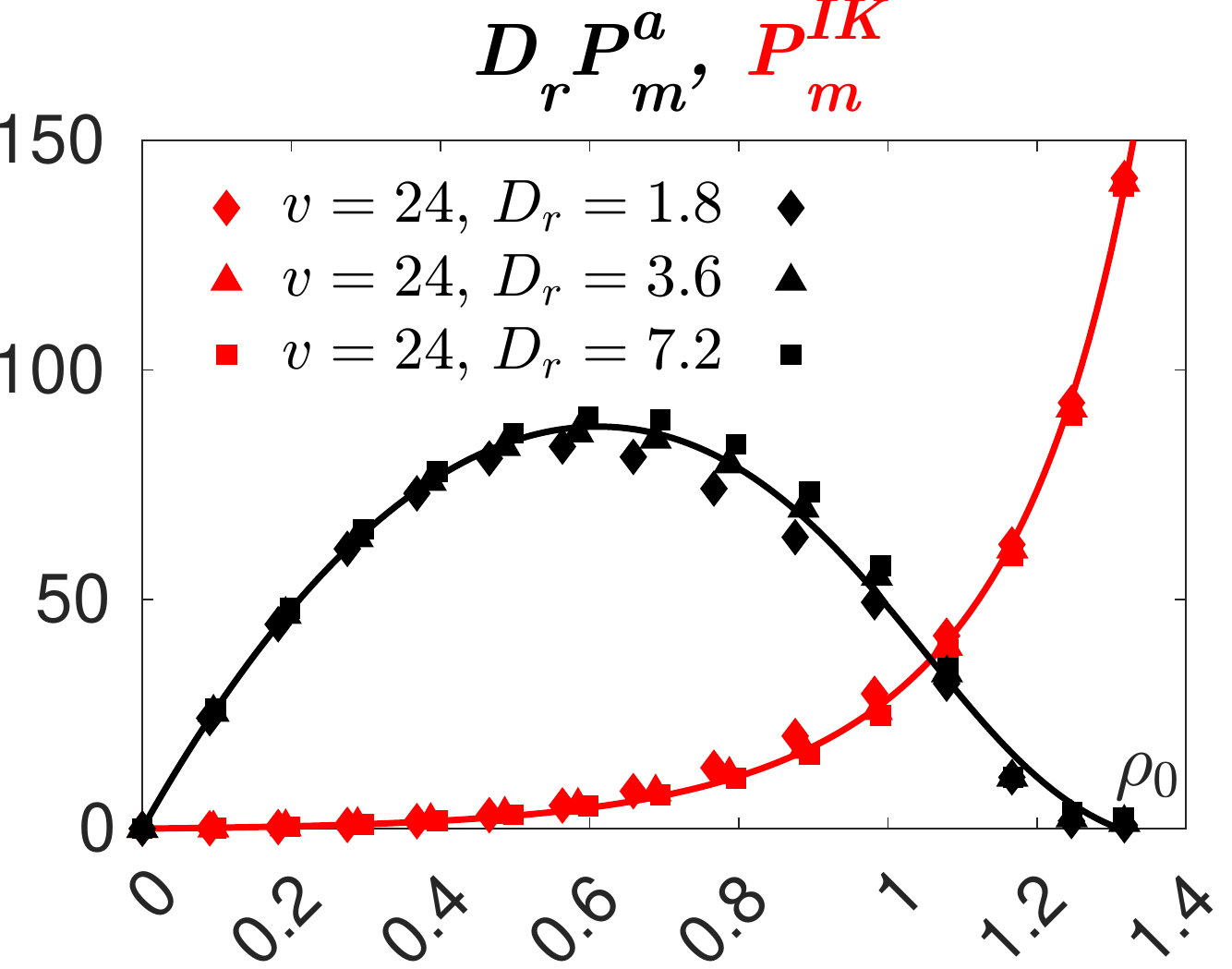}};
    \path (10.7cm,0) node {\includegraphics[height=4cm]{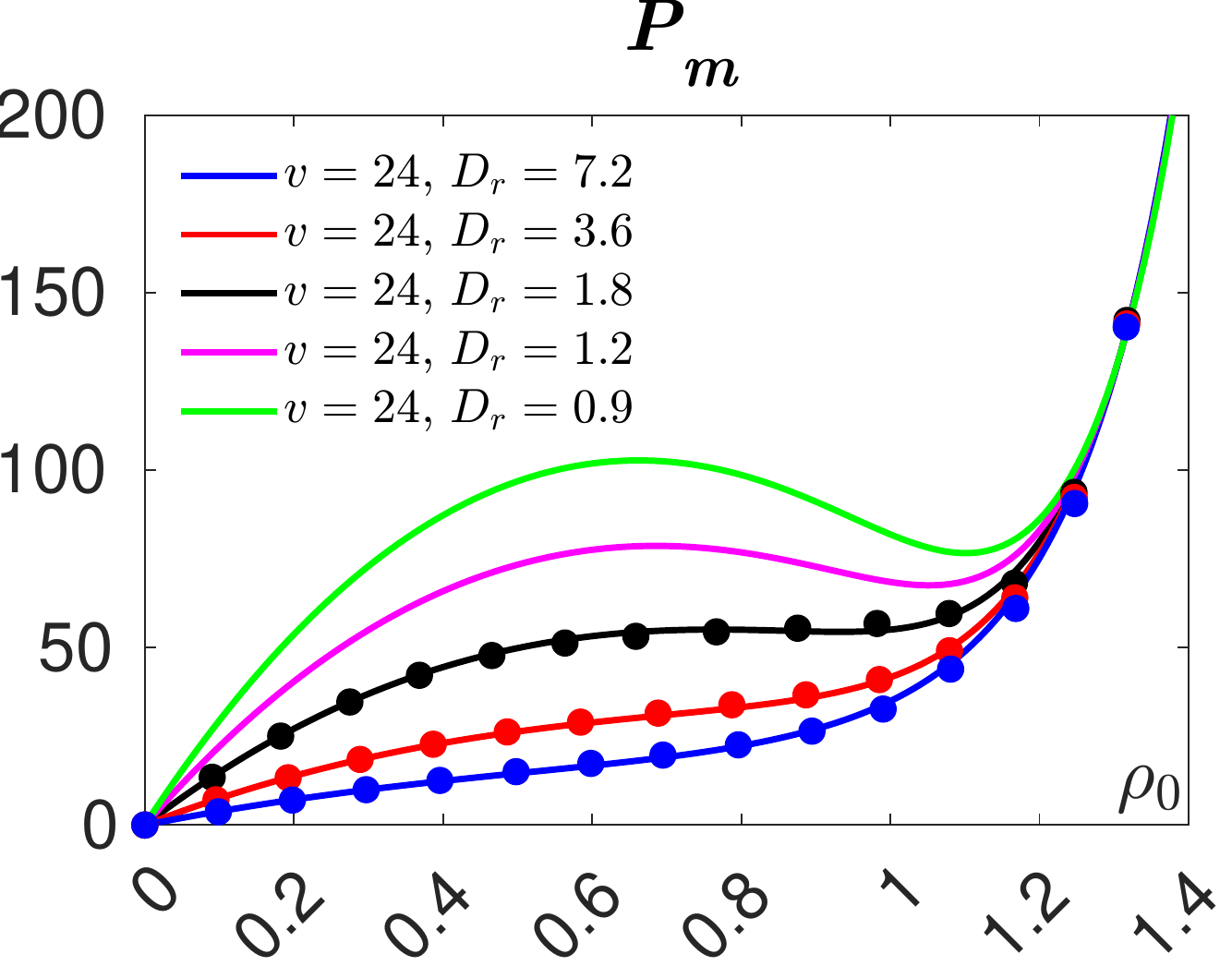}};
    
    \node at (2cm,2.2cm) {\textbf{(a)}};
    \node at (7.35cm,2.2cm) {\textbf{(b)}};
    \node at (12.7cm,2.2cm) {\textbf{(c)}};
  \end{tikzpicture}}\fi
  \includegraphics[width=\textwidth]{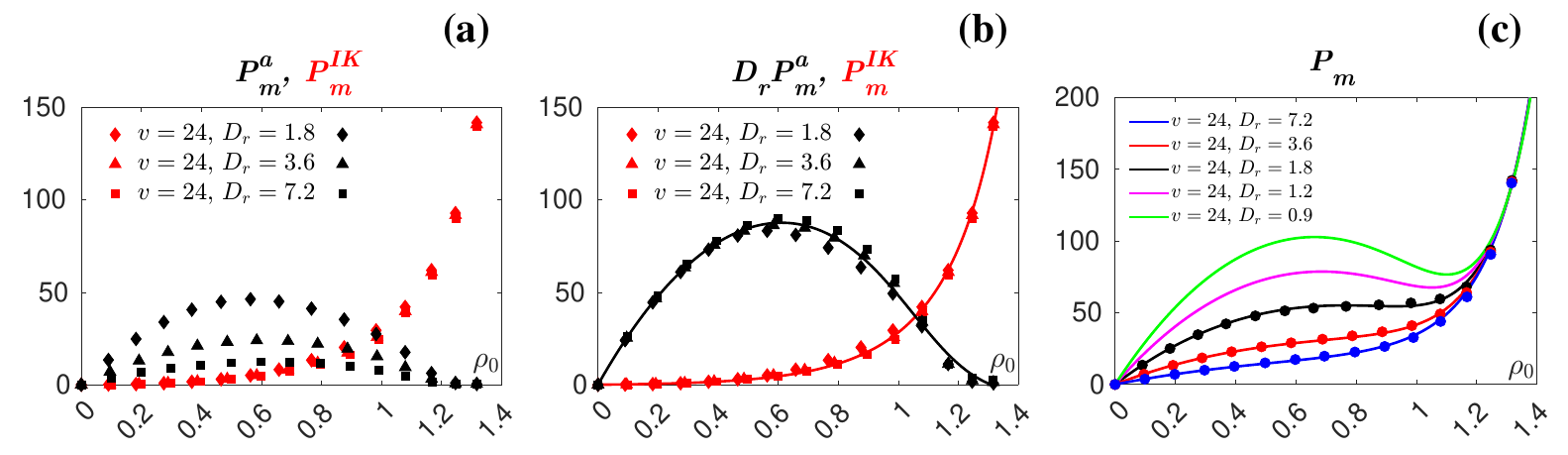}
    \caption{Pressure of PFAPs interacting via a repulsive WCA potential. (See~\cite{solon2018njp} for numerical details.) \textbf{(a)} $P_m^a$ and $P_m^{IK}$ as the average density $\rho_0$ is varied, measured for values of $D_r$ and $v_0$ leading to homogeneous systems. \textbf{(b)} Rescaling the pressures as $D_r P_m^a$ and $P_m^{IK}$ shows the scaling~\eqref{eq:PmIK} and~\eqref{eq:Pma} of the pressures as $D_r$ varies to be satisfied. The solid lines show fits to the simulation data, as detailed in~\cite{solon2018njp}. \textbf{(c)} The total pressure $P_m=P_m^a+P_m^{IK}$ as a function of $\rho_0$. The curves for $D_r=1.2$ and $D_r=0.9$ are obtained by extrapolating the fitted curves in panel b), using the scaling laws predicted in Eqs.~\eqref{eq:PmIK} and~\eqref{eq:Pma}. At a sufficiently small $D_r$, and hence a sufficiently large persistence length, the decreasing trend of $P_m^a$ for $\rho_0\geq \rho^*/2$ overcomes the increasing trend of $P_m^{IK}$. This leads to a total decreasing pressure, $P_m'(\rho_0)<0$, and hence a linear instability.}
    \label{fig:PFAPPressure}
\end{figure}
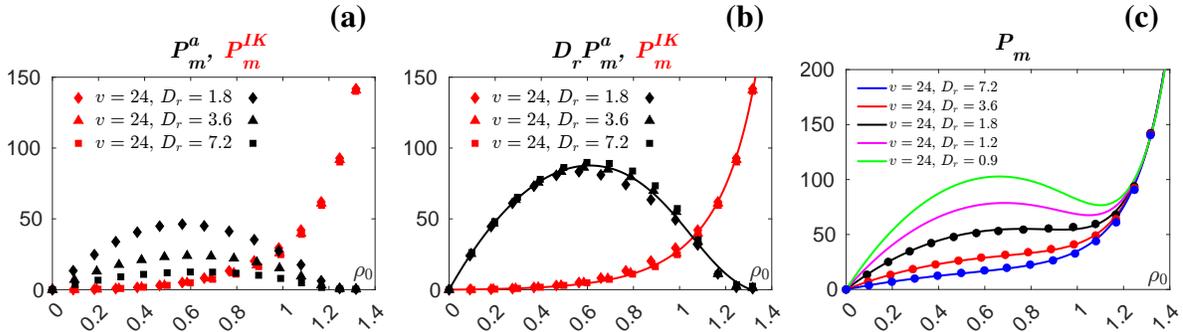

We are now ready to account for the shape of the phase diagram of
PFAPs. At small $v_0$, the active pressure, which scales as $v_0^2$,
is negligible. The stabilizing effect of repulsive interactions
dominate and, like in passive systems, they promote a homogeneous
phase. At higher speeds, however, the active pressure becomes
comparable to the Irving-Kirkwood contribution. When $\rho_0\geq
\rho^*/2$, the decreasing trend of $P_m^a$ may thus overcome the
increasing trend of $P_m^{IK}$, leading to an overall decreasing
pressure. At close packing, the divergence of $P_m^{IK}$ stabilizes
again the system. Balancing Eqs.~\eqref{eq:PmIK} and~\eqref{eq:Pma}
shows that the transition between the two regimes is controlled by the
dimensionless number $v_0/(D_r r_0)$ which compares the particle
persistence length to the particle size. In the presence of a finite
translational diffusivity $D_t$, when $D_r$ and $D_t$ are related by
Stokes relations, this criterion can be expressed in terms of a
P\'eclet number $v_0 r_0/D_t$. However appealing, this choice is
unfortunate since the phase diagram of MIPS is barely altered by
sending $D_t$ to zero while maintaining $D_r$ constant and the physics
described above has nothing to do with translational diffusion. The
nondimensional persistence length $v_0/(D_r r_0)$ is thus a better observable to characterize the transition. The scaling
laws~\eqref{eq:PmIK} and~\eqref{eq:Pma} for the `passive' and `active'
contributions to the pressure can be tested by measuring $P_m^a$ and
$P_m^{IK}$ in homogeneous systems at density $\rho_0$. We report these
measurements in Fig.~\ref{fig:PFAPPressure} and use them to
extrapolate the behaviour of the pressure at larger persistence
lengths. Above a `critical' persistence length, $P_m(\rho_0)$ is
non-monotonous, which allows defining a spinodal region where
$P_m'(\rho_0)<0$ and where a mean-field instability is predicted.

\begin{figure}
    \centering
    \includegraphics[totalheight=4cm]{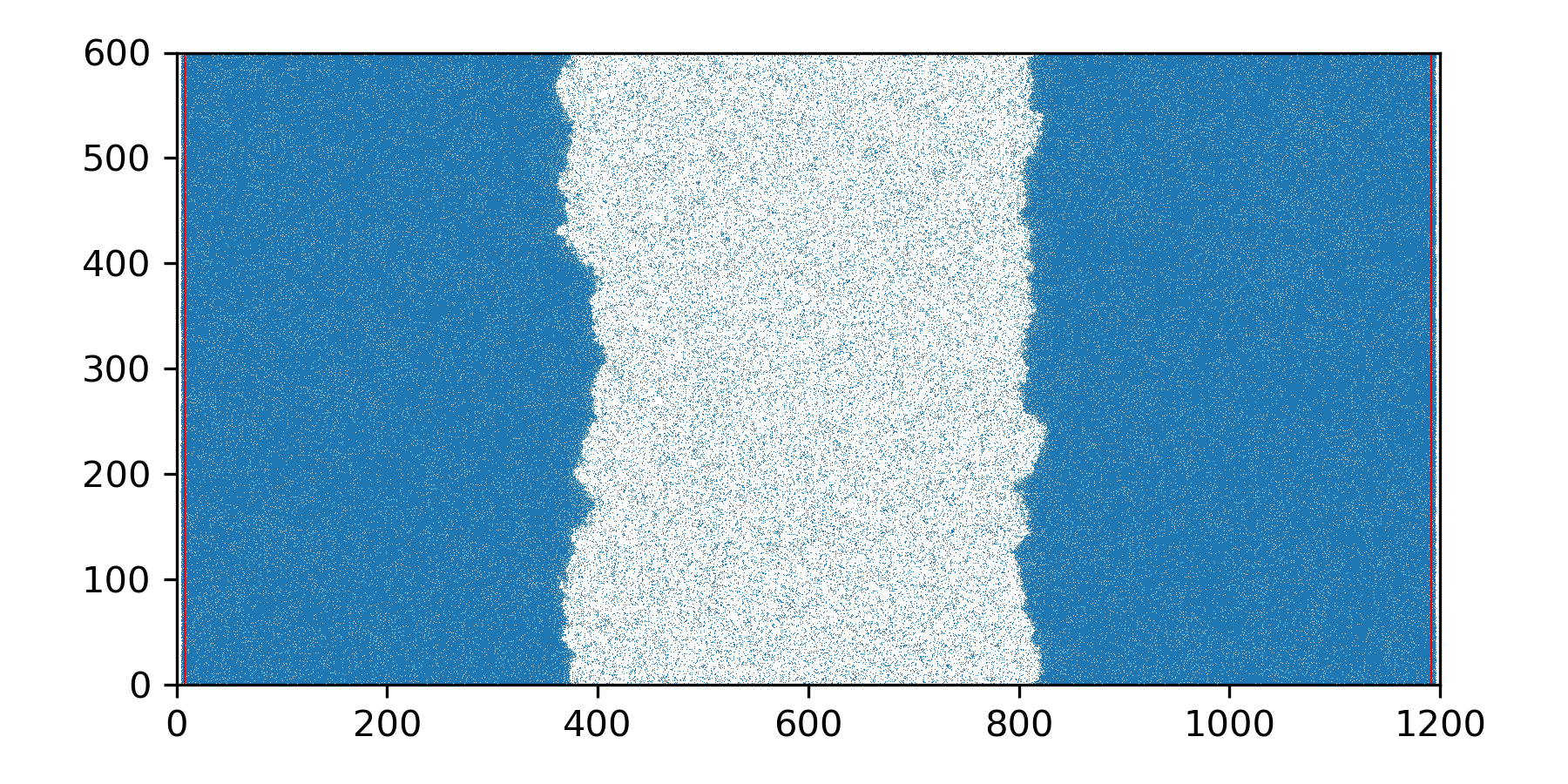}\includegraphics[totalheight=4cm]{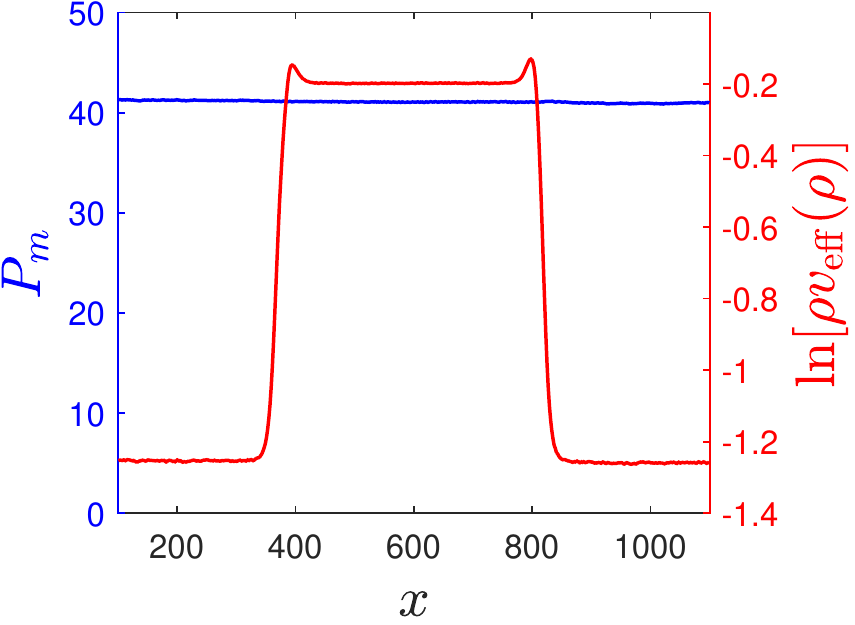}
    \caption{{\bf Left:} Self-propelled ABPs interacting via a harmonic repulsive potential and undergoing MIPS. The particles are confined by hard walls at $x=8$ and $x=1192$ while they experience periodic boundary conditions along the $y$ direction. {\bf Right:} We measure the pressure $P_m$ and the chemical potential $\mu=\ln[\rho v_{\rm eff}(\rho)]$ in slabs of width $\Delta x=2$. The mechanical pressure, which corresponds to $-\sigma_{xx}$ is clearly uniform accross the system whereas the chemical potential that would be defined using the QSAPs approximation is not.}
    \label{fig:Pvsmu}
\end{figure}

Note that the path laid out above not only explains the instability
leading to MIPS for PFAPs, but also how it overcomes the stabilizing
effect of repulsive forces, something which could not be discussed
within the QSAPs framework. Finally, note that we have obtained a
second important result: since $\bfJ=\mathfrak{u} \nabla \cdot\sigma$,
stress tensor and pressure are homogeneous in flux-free steady
states. The mechanical pressure thus acts as a state variable in the
MIPS exhibited by PFAPs, and it is thus equal in coexisting
phases. Figure~\ref{fig:Pvsmu} shows an assembly of PFAPs undergoing
MIPS. The right panel shows measurements of $P_m$ and of a putative
chemical potential defined as $\mu\equiv \ln[\rho v_{\rm eff}(\rho)]$
throughout the system. While $P_m$ is, as predicted, uniform, $\mu$ is
very different between the two phases.

Before concluding this chapter, we finally discuss in the next section
the MIPS observed in lattice models.

\section{MIPS in lattice models}
\label{sec:lattice}
\subsection{Models and phenomenology}

Despite being often harder to connect to experimental systems, lattice
gases are workhorse models in statistical physics because they allow
both for analytical progress and for faster simulations than their
off-lattice counterparts. Lattice-based models have been used to model
active systems in a variety of settings, in particular to study the
transition to collective
motion~\cite{csahok1995lattice,o1999alternating,peruani2011traffic,solon2013revisiting}
and motility-induced phase
separation~\cite{thompson2011lattice,soto2014run,manacorda2017lattice,whitelam2018phase,kourbane2018exact,shi2020self,ro2021disorder}. Here,
we follow~\cite{thompson2011lattice} and consider the run-and-tumble
dynamics of $N$ particles on a two-dimensional lattice. Particle $k$
occupies a site $\bfi_k\equiv(i_k,j_k)$ and carries an orientation
vector $\bfu_k\in\{(\pm 1,0),(0,\pm 1)\}$. In the non-interacting
case, the particle hops with a constant rate $p$ from site $\bfi_k$ to
site $\bfi_k+\bfu_k$ and fully randomizes its orientation at rate
$\alpha$. Denoting $a$ the lattice spacing, this leads to a
`propulsion speed' $v=pa$, a persistence length $\ell_p=p a / \alpha$
and to a large-scale effective diffusivity~(See
appendix~\ref{app:Diff}):
\begin{equation}\label{eq:difflattice}
  D_0=\frac{p^2 a^2}{2\alpha}\left(1+\frac \alpha {2p}\right)\;.
\end{equation}
In the limit $\ell_p\gg a$, we thus recover the large-scale diffusivity of
an off-lattice RTP in 2D, $D_0=v^2/(2\alpha)$.

To model repulsive forces and steric interactions,  the
hopping rule can be replaced by a partial exclusion
process~\cite{schutz1994non,tailleur2008mapping}: a particle hops from site $\bfi_k$ to
site $\bfi_k+\bfu_k$ with a rate
\begin{equation}\label{eq:vofrholattice}
  W(\bfi_k\to \bfi_k+\bfu_k) = p \left(1-\frac{n_{\bfi_k+\bfu_k}}{n_{\rm m}}\right) \Theta(n_{\rm m}-n_{\bfi_k+\bfu_k})\;,
\end{equation}
where $n_{\bf j}$ is the total number of particles on site ${\bf j}$
and $n_{\rm m}$ the maximal occupancy. Note that this amounts to
making self-propulsion depend on density, which thus leads to a
discretized version of quorum-sensing interactions. The linear
decay~\eqref{eq:vofrholattice} is, however, reminiscent of the
effective self-propulsion speed~\eqref{eq:PFAPSeffspeed} measured in
the presence of pairwise forces. MIPS has been reported for such
lattice models at large enough
$\ell_p$~\cite{tailleur2008mapping}. Interestingly, for the very model
described above, MIPS also requires $n_{\rm m}>1$~\cite{soto2014run},
a constraint which is released if translational noise is added to the
model (for instance in the form of a non-vanishing symmetric
transverse hopping rate~\cite{whitelam2018phase}).

\begin{figure}
    \centering
    \begin{tikzpicture}
    \path (0,0) node {\includegraphics[height=5cm]{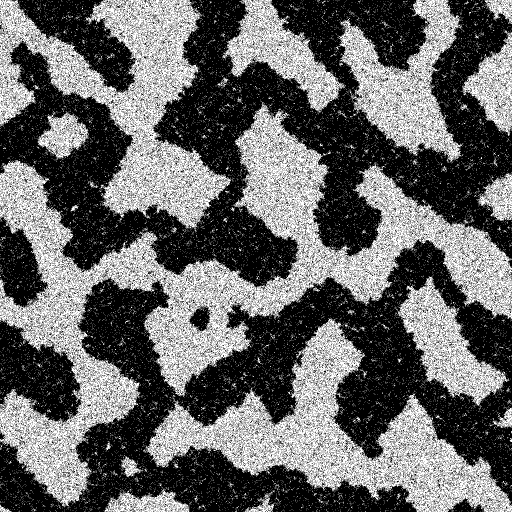}};
    \path (11cm,0) node {\includegraphics[height=5cm]{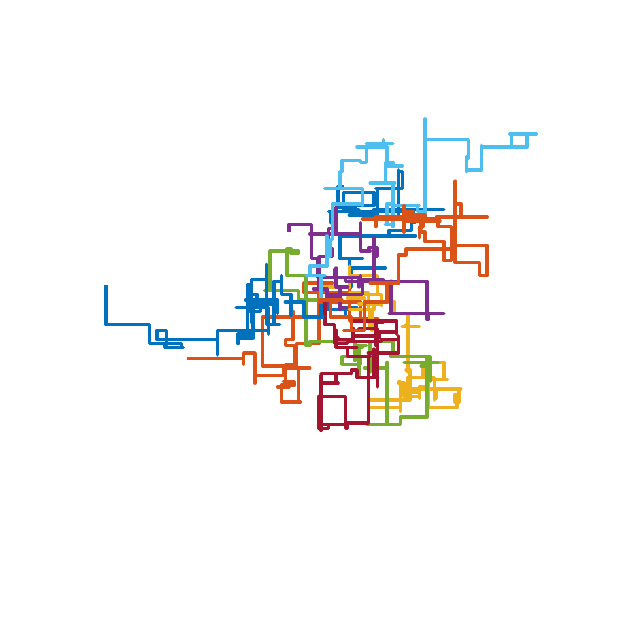}};
    \path (6cm,0) node {\includegraphics[height=5cm]{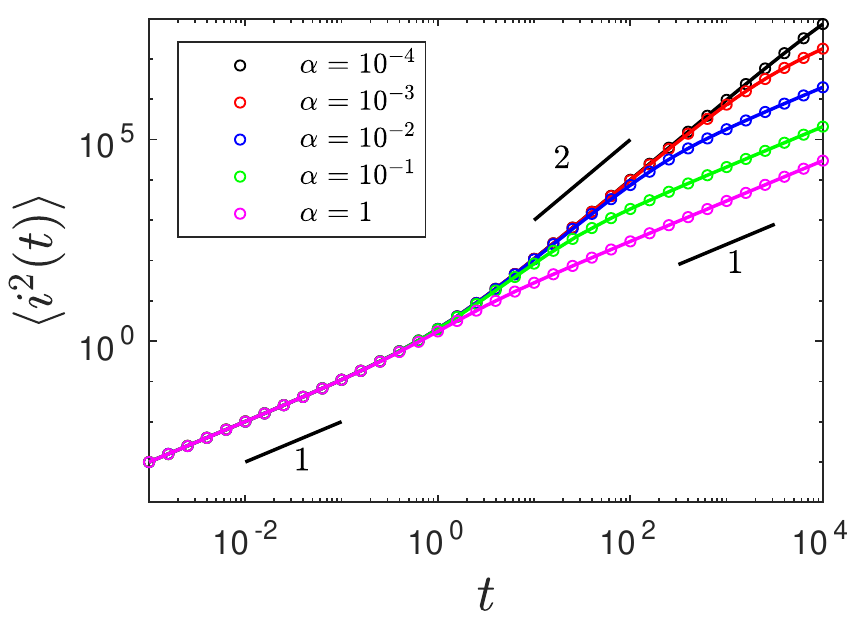}};
    
    \node at (2cm,2.8cm) {\textbf{(a)}};
    \node at (8.8cm,2.8cm) {\textbf{(b)}};
    \node at (12.5cm,2.8cm) {\textbf{(c)}};
    \end{tikzpicture}
    \caption{MIPS on lattice. \textbf{(a)} A snapshot at large time showing the phase separation of RTPs on a 2D square lattice of $512\times 512$ sites. $p=10$, $\alpha=1$, $n_{\rm m}=5$. The initial density is $\langle n\rangle=3$. \textbf{(b)} Mean squared displacement of a free RTP on a 2D square lattice. Solid lines are from Eq. \eqref{eq:MSDlatticeRTP} and circles are from simulations. $p=1$. \textbf{(c)} Ten trajectories of non-interacting RTPs starting from the same position on a 2D lattice. $p=10$, $\alpha=1$.}
    \label{fig:latticeMIPS}
\end{figure}
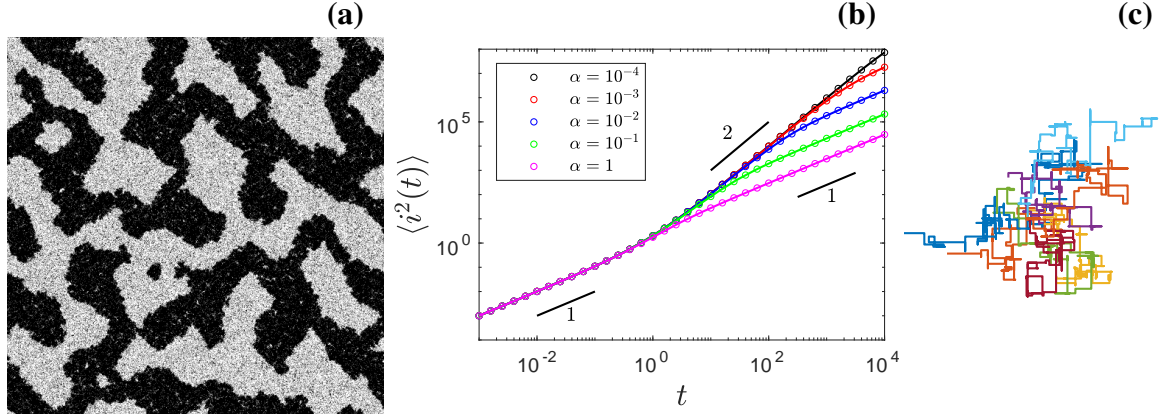

\subsection{Linear instability at mean-field level}

To illustrate the power of lattice approaches, we now derive the
criterion for linear instability leading to MIPS. To lighten the
notations, we consider a one-dimensional lattice. In this case, the
linear instability leads to alternating finite-size domains of gas and
liquid~\cite{tailleur_statistical_2008}. The physics of the
instability is, however, identical to the one leading to MIPS in
higher dimensions.

We first remind the generic expression of the master equation that
governs the time evolution of the probability $P(\C,t)$ to observe a
configuration $\C$ at time $t$:
\begin{equation}\label{eq:ME}
  \partial_t P(\C,t)=\sum_{\C'} W(\C'\to\C) P(\C',t)-W(\C\to\C') P(\C,t)\;,
\end{equation}
where $W(\C\to \C')$ is the rate at which the system transitions from
configuration $\C$ to configuration $\C'$.  Consider a generic
observable $O(\C)$. Its average value at time $t$ is defined as
\begin{equation}
  \langle O (t)\rangle \equiv \sum_{\C} O(\C) P(\C,t)\;.
\end{equation}
Using the master equation~\eqref{eq:ME}, the time evolution of $
\langle O (t)\rangle$ can then be written as
\begin{equation}\label{eq:dynavobs}
  \frac{d}{dt}   \langle O (t)\rangle=\sum_{\C} \Big\{ \sum_{\C'}[O(\C')-O(\C)] W (\C\to\C')
  \Big \}P(\C,t)=\Big\langle \sum_{\C'}[O(\C')-O(\C)] W (\C\to\C') \Big\rangle_{\C}
\end{equation}
The evolution of the mean of the observable is thus given by its
average variation during hops out of configuration $\C$ weighted by
the corresponding rates.

Let us denote $n_i^\pm$ the numbers of particles at site $i$ whose
orientations are $\pm e_x$, respectively. Applying Eq.~\eqref{eq:dynavobs} to
$\langle n_i^+\rangle$ and $\langle n_i^-\rangle$ leads to
\begin{eqnarray}
  \partial_t \langle n_i^+(t)\rangle &=& \Big\langle p n_{i-1}^+ \Big(1-\frac {n_i}{n_{\rm m}}\Big)-p n_{i}^+ \Big(1-\frac {n_{i+1}}{n_{\rm m}}\Big)+\frac{\alpha}{2} n_i^- - \frac{\alpha}{2} n_i^+\Big\rangle\\
  \partial_t \langle n_i^-(t)\rangle &=& \Big\langle p n_{i+1}^- \Big(1-\frac {n_i}{n_{\rm m}}\Big)-p n_{i}^- \Big(1-\frac {n_{i-1}}{n_{\rm m}}\Big)-\frac{\alpha}{2} n_i^- + \frac{\alpha}{2} n_i^+\Big\rangle
\end{eqnarray}
At this stage, the dynamics of the one-point functions
$\rho_i^\pm(t)\equiv \langle n_i^\pm(t)\rangle$ involve two-point
functions, such as $\langle n_i^\pm n_{i+1}^\pm\rangle$, and are thus not
closed. A popular choice is then to take a ``mean-field''
approximation that amounts to neglecting all correlations and to
factorize all averages~\cite{blythe2007nonequilibrium}. Under this
approximation, the dynamics of $\rho_i^\pm(t)$ are closed and given by
\begin{equation}\label{eq:dynrhoi}
  \partial_t \rho_i^\pm =  p \rho_{i\mp 1}^\pm \Big(1-\frac {\rho_i}{n_{\rm m}}\Big)-p \rho_{i}^\pm \Big(1-\frac {\rho_{i\pm 1}}{n_{\rm m}}\Big)\pm \frac{\alpha}{2} \rho_i^- \mp \frac{\alpha}{2} \rho_i^+
\end{equation}

To make progress, we assume that $\rho_i^+$ and $\rho_i^-$ vary on
scales much larger than the lattice spacing $a$. (Note that, while
this is reasonnable when $\ell_p\gg a$, the interactions may very well
generate sharp fronts that would not satisfy this hypothesis.) We then
introduce the density fields of left- and right-going particles,
denoted $\rho^-(x)$ and $\rho^+(x)$, respectively, that interpolate
the $\rho_i^\pm$ and satisfy $\rho_i^\pm=a\rho^\pm(x=i a)$. Assuming
$\rho^\pm(x)$ to vary smoothly, we expand $  \rho_{i\pm 1}^\pm$ as
\begin{equation}
  \rho_{i\pm 1}^\pm= a\rho^\pm(x) \pm a^2 \partial_x \rho^\pm(x) + \frac{a^3}{2} \partial_{xx} \rho^\pm(x)\;.
\end{equation}
We now introduce $\rho_{\rm m}=n_{\rm m}/a$,
$\rho(x)=\rho^+(x)+\rho^-(x)$ and
$m(x)=\rho^+(x)-\rho^-(x)$. Taylor-expanding Eq.~\eqref{eq:dynrhoi} to
second order in gradients yields the following dynamics for the fields $\rho(x,t)$
and $m(x,t)$:
\begin{align}
  \partial_t \rho (x)&=-\partial_x \Big[p a m(x) \Big(1-\frac{\rho(x)}{\rho_{\rm m}}\Big)-\frac{p a^2}2 \partial_x \rho(x)\Big]\label{eq:dynrho1}\\
  \partial_t m (x)&=-\alpha m(x) -\partial_x \Big[p a \rho(x) \Big(1-\frac{\rho(x)}{\rho_{\rm m}}\Big)\Big]+\frac{p a^2}2 \Big[ \partial_{xx} m(x)\Big(1-\frac{\rho(x)}{\rho_{\rm m}}\Big)+ \frac{m\partial_{xx} \rho(x)}{\rho_{\rm m}}\Big]\;,\label{eq:dynm1}
\end{align}
where the time-dependence of the fields has been ommitted for
concision. We note the presence of a `bare' diffusivity $pa^2/2$ in
the dynamics of $\rho$ which is nothing but the correction to the
large-scale diffusivity~\eqref{eq:difflattice} due to the discreteness
of the lattice.

Let us now consider the limit in which $v=pa$ is kept fixed but the
lattice spacing $a$ is assumed to be very small. The persistence
length $\ell_p=v/\alpha$ is thus large compared to $a$ and we hope the
irregularities due to the lattice to be negligible, so that the model
should be as close as possible to a continuous one. Assuming the
gradients of $\rho$ and $m$ to remain finite, the terms proportional
to $pa^2/2$ are negligible in the small-$a$ limit. The
dynamics~\eqref{eq:dynrho1} and~\eqref{eq:dynm1} thus reduce to
\begin{align}
  \partial_t \rho (x)&=-\partial_x \Big[v m(x) \Big(1-\frac{\rho(x)}{\rho_{\rm m}}\Big)\Big]\label{eq:dynrho2}\\
  \partial_t m (x)&=-\alpha m(x) -\partial_x \Big[v \rho(x) \Big(1-\frac{\rho(x)}{\rho_{\rm m}}\Big)\Big]\;.\label{eq:dynm2}
\end{align}
As in the Section~\eqref{sec:vofrFP}, $\rho(x)$ is a conserved field
whose relaxation time diverges with the system size. On the contrary,
the field $m(x)$ relaxes on a time-scale of order ${\cal
  O}(\alpha^{-1})$ towards
\begin{align}
  m (x)&= -\partial_x \Big[\frac{v}\alpha \rho(x) \Big(1-\frac{\rho(x)}{\rho_{\rm m}}\Big)\Big]\;.\label{eq:dynm3}
\end{align}
It then follows adiabatically the evolution of $\rho(x,t)$ on
hydrodynamic time scales according to Eq.~\eqref{eq:dynm3}. All in
all, this yields the following non-linear diffusion equation for
$\rho(x,t)$
\begin{equation}\label{eq:nonlineardiff}
  \partial_t\rho(x,t)=\partial_x[D_{\rm coll}(\rho(x,t))\partial_x\rho(x,t)]\quad\text{where}\quad D_{\rm coll}=\frac{v^2}{\alpha} \left(1-\frac{\rho(x)}{\rho_{\rm m}}\right) \left(1-\frac{2\rho(x)}{\rho_{\rm m}}\right)
\end{equation}
is an effective, collective diffusivity.

The stability of a homogeneous profile at density $\rho_0$ is now
simple to predict: when $D_{\rm coll}(\rho_0)>0$, fluctuations are
damped and the homogeneous phase is linearly stable. All our
approximations are self-consistent and the
dynamics~\eqref{eq:nonlineardiff} would faithfully describe the
large-scale relaxation of the average density field towards $\rho_0$. When $D_{\rm
  coll}(\rho_0)<0$, on the contrary, fluctuations are amplified and
drive a linear instability. In this case, sharp profiles are expected
and the gradient expansion should be extended to higher order to
compute the inhomogeneous profiles.

Interestingly, the instability {criterion} amounts to
$\rho_0>\rho_m/2$. Considering that the exclusion rule leads to an
effective ``velocity'' $v(\rho)=v (1-\rho/\rho_m)$, this {criterion}
is equivalent to $v'(\rho)/v(\rho)<-1/\rho$. In the limit $\ell_p\gg
a$, the lattice gas thus gives back, using a much simpler computation,
the linear instability {criterion}~\eqref{eq:spinodal} for QSAPs that
we derived in Section~\ref{sec:local}.

Progress beyond the linear instability is possible using
field-theoretical methods that allow constructing the fluctuating
hydrodynamics of lattice gases~\cite{thompson2011lattice}. As for
QSAPs, a mapping to an effective equilibrium dynamics can be
constructed and the phase diagram can be predicted, using methods
identical to those presented in Section~\ref{sec:generalized
  thermo}. Furthermore, exact results can be obtained on the phase
diagrams of lattice models~\cite{kourbane2018exact} using recent
results from the mathematical-physics
literature~\cite{erignoux2018hydrodynamic}.  These results are,
however, beyond the scope of this Chapter.

Having reviewed the three most important classes of systems in which
MIPS has been reported, we now conclude this Chapter with a discussion
of what is \textit{not} known about MIPS and list some current
exciting topics of research.

\section{Conclusion}
\label{sec:conclu}

Motility-induced phase separation is arguably the simplest non-trivial collective phenomenon experienced by active particles. It is non-trivial because its microscopic roots are very different from what leads to phase separation in equilibrium, and because we do not have any theoretical tools that can \textit{a priori} account for its phenomenology. On the other hand, the sole hydrodynamic field is the density, which makes the system particularly simple to describe at the coarse-grained level. We have here focused on the most established properties of MIPS: the microscopic instability that drives phase separation and the coexisting densities in macroscopic phase-separated systems. Many questions regarding MIPS physics and its interplay with other ingredients frequently encountered in active systems however remain open.

In the simplest setting of scalar active matter, the question of the coarsening dynamics leading to MIPS would deserve a thorough investigation. In particular, some active-matter systems have been shown to admit a negative surface {tension}~\cite{bialke2015negative,solon2018njp}. At the coarse-grained level, this has been predicted to arrest MIPS or to lead to a coexistence between a gaseous phase and a bubbly liquid~\cite{tjhung2018cluster}. The latter has been observed in microscopic simulations of ABPs and other related models~\cite{caporusso2020motility,shi2020self} {although} a clear connection between microscopic and macroscopic models is still missing. 

Then, the simple models discussed in this review lack many ingredients frequently found in active systems. From phoretic~\cite{pohl2014dynamic} and hydrodynamic~\cite{matas-navarro_hydrodynamic_2014} interactions to population dynamics~\cite{cates_arrested_2010,liu_sequential_2011,grafke2017spatiotemporal} they can lead to strong alterations of MIPS physics. Stepping out of the realm of scalar active matter, the interplay between MIPS and aligning interactions is rich~\cite{peruani2011traffic,farrell2012pattern,sese2018velocity,shi2018self,geyer_freezing_2019} and calls for deeper {investigations}.

To finish on a positive note, let us stress that the exact mapping onto passive colloids at the fluctuating-hydrodynamic level discussed in Section~\ref{sec:nonlocal} offers an interesting perspective. It indeed suggests that MIPS could be of one the many collective behaviors of active systems that could resemble equilibrium physics but that would emerge from microscopic interactions of a very different nature from their equilibrium counterparts. Whether this offers a new route to account for phenomena observed in biological systems or a pathway to engineer active materials is an exciting perspective that calls for further studies. 

\vspace{1cm}

\noindent\textit{Acknowledgments.} We thank Alberto Dinelli, Luigi Gentile, Yariv Kafri, Christina Kurtzhaler, Howard Stone, and Sunghan Ro for their usefuls comments on this chapter.

\bibliography{biblio.bib}
\bibliographystyle{rsc}

\appendix{}

\section{Mean squared displacement of an RTP on lattice}
\label{app:Diff}
\subsection{Diffusivity}
Let us first provide a simple derivation of the large-scale diffusivity \eqref{eq:difflattice}. We consider a single run-and-tumble particle on a $d$-dimensional square lattice. In the unit of lattice spacing $a$, we denote the displacement of the $j$-th run as $\mathbf{s}_j$. The total displacement $\Delta\bfi$ at time $t$ is, by definition,
\begin{equation}
    \Delta\bfi(t) = \sum_{j=1}^{n(t)}\mathbf{s}_j\;,
\end{equation}
where $n(t)$ is the number of runs until time $t$. The {mean squared} displacement (MSD) at time $t$ is then given by
\begin{equation}\label{eq:MSD}
 \left\langle \Delta i^2(t)\right\rangle=\left\langle\sum_{j=1}^{n(t)}s^2_i\right\rangle+2\left\langle \sum_{j<k}\mathbf{s}_j\cdot\mathbf{s}_k\right\rangle\;.
\end{equation}
Note that both $n(t)$ and the $\mathbf{s}_i$'s are random variables. 

Since we consider isotropic tumbles, $\langle\mathbf{s}_j\rangle=0$ and the displacements during different runs are independent so that, for $i\neq j$,
\begin{equation}
    \left\langle\mathbf{s}_i\cdot\mathbf{s}_j\right\rangle=\langle\mathbf{s}_i\rangle\cdot\langle\mathbf{s}_j\rangle=0\;,
\end{equation}
and the last term in~\eqref{eq:MSD} vanishes.

Since the run lengths are identically distributed, Wald's identity states that, in the large $t$ limit,
\begin{equation}
    \left\langle \Delta i^2(t)\right\rangle\underset{t\to\infty}{\sim}\left\langle n(t)\right\rangle\left\langle s^2\right\rangle\;,
\end{equation}
where $\langle s^2 \rangle$ is the second moment of the run length (measured in units of the lattice spacing). The average number of runs is given by $\langle n(t)\rangle =\alpha t$. 
Every time the particle changes configuration, it has a probability $p/(p+\alpha)$ to hop and a probability $\alpha/(p+\alpha)$ to tumble. The probability distribution $P(s)$ that a run corresponds to $s$ hops is thus given by
\begin{equation}\label{eq:pofs}
    P(s)=\frac{p^s\alpha}{(p+\alpha)^{s+1}}\;.
\end{equation}
Note that the very last run is not necessarily finished at time t, and is thus not distributed according to Eq. (105). This is why this heuristic calculation applies only in the large-time limit. By direct calculation, Eq.~\eqref{eq:pofs} leads to
\begin{equation}
    \left\langle s^2\right\rangle=\sum_{s=1}^{\infty}P(s)s^2=\frac{2p^2+\alpha p}{\alpha^2}\;. \label{eq:means2lattice}
\end{equation}
All in all, we thus find that the late-time MSD satisfies
\begin{equation}
    \left\langle \Delta i^2(t)\right\rangle\underset{t\to\infty}{\sim}\left\langle n(t)\right\rangle\left\langle s^2\right\rangle=\frac{2p^2+\alpha p}{\alpha}t\;. \label{eq:MSDlatticeRTPlarget}
\end{equation}
Restoring the unit of length, the diffusivity satisfies $\left\langle a^2 \Delta i^2(t)\right\rangle  \sim 2d D_0 t$ in the large-$t$ limit so that
\begin{equation}
    D_0=\frac{2p^2+\alpha p}{2d\alpha}a^2=\frac{p^2a^2}{d\alpha}\left(1+\frac{\alpha}{2p}\right)\;.
\end{equation}
Setting $d=2$ for a 2D lattice leads to Eq.~\eqref{eq:difflattice}.

This heuristic derivation provides an insight into the large-time asymptotic behaviour of the MSD. We note that the term $\alpha/(2p)$ in the bracket, which is absent in the continuous model, stems from the non-exponential statistics of the run length $s$. In the continuous model, $s$ is exponentially distributed so that $\langle s^2\rangle=2\langle s\rangle^2$. This is not the case on lattice for finite $p$ and $\alpha$, but Eq.~\eqref{eq:pofs} shows that, as $p\gg \alpha$, $P(s)\sim \frac\alpha p \exp[-\alpha s/p]$, which thus leads back to the result of the continuous model in this limit. 

\subsection{Mean-square displacement at all times}

We now provide the full expression of the MSD at all times. We consider the probability density $P(\bfi,\bfu,t)$ of finding the particle at site $\bfi$ with orientation $\bfu$ at time $t$, where $\bfi=(i_1,i_2,\cdots,i_d)$ and $\bfu$ is one of the $2d$ unit vectors $\bfu^{(\pm j)}$ with $j\in \llbracket 1;d\rrbracket$ and $u^{(\pm j)}_k \equiv\pm \delta_{jk}$. Since the system is isotropic, we can, without loss of generality, consider only the displacements along the direction $i_1$. We denote the corresponding marginal probability densities $P_\pm(i_1,t)=\sum_{i_2,\cdots,i_d}P(\bfi,\bfu^{(\pm 1)},t)$ and $P_0(i_1,t)=\sum_{i_2,\cdots,i_d}\sum_{\bfu\neq \bfu^{(\pm 1)}}P(\bfi,\bfu,t)$. $P_+(i_1,t)$ and $P_-(i_1,t)$ correspond to the particle being in a site $\bfi$ whose first component equals $i_1$, going towards greater and smaller values of $i_1$, respectively. $P_0$ correspond to the particle going along one of the $d-1$ orthogonal directions. The master equations can be written as
\begin{align}
    \frac{\partial }{\partial t}P_\pm(i_1,t)=&pP_\pm(i_1\mp 1,t)-pP_\pm(i_1,t)-\alpha P_\pm(i_1,t)+\frac{\alpha}{2d}[P_+(i_1,t)+P_-(i_1,t)+P_0(i_1,t)]\;, \label{eq:mastereqlatticex1}\\
    \frac{\partial }{\partial t}P_0(i_1,t)=&-\alpha P_0(i_1,t)+\alpha \frac{2d-2}{2d}[P_+(i_1,t)+P_-(i_1,t)+P_0(i_1,t)]\;. \label{eq:mastereqlattice0}
\end{align}

We introduce the generating functions
\begin{equation}
    G_{+,-,0}(z,t)=\sum_{i_1=-\infty}^{\infty} z^{i_1}P_{+,-,0}(i_1,t)\quad\text{and}\quad G(z,t)=G_++G_-+G_0\;.
\end{equation}
Direct computation shows that $G'(z=1,t)=\langle i_1(t)\rangle$ and $G''(z=1,t)=\langle i_1^2(t)\rangle-\langle i_1(t)\rangle$, where the prime denotes a derivative with respect to $z$. We consider the initial condition $i_1(t=0)=0$ so that the MSD can be obtained from the derivatives of $G$ as
\begin{equation}\label{eq:MSDandG}
    \langle\Delta i_1^2(t)\rangle=G''(1,t)+G'(1,t)
\end{equation}

Multiplying Eqs. \eqref{eq:mastereqlatticex1} and \eqref{eq:mastereqlattice0} by $z^{i_1}$ and summing on $i_1$, we find that
\begin{align}
    \frac{\partial}{\partial t}G_+(z,t)=&p(z-1)G_+-\alpha G_++\frac{\alpha}{2d}G\;, \label{eq:gflatticegp}\\
    \frac{\partial}{\partial t}G_-(z,t)=&p\left(\frac{1}{z}-1\right)G_--\alpha G_-+\frac{\alpha}{2d}G\;, \label{eq:gflatticegm}\\
    \frac{\partial}{\partial t}G_0(z,t)=&-\alpha G_0+\alpha\frac{(2d-2)}{2d}G\;, \label{eq:gflatticeg0}
\end{align}
which form a closed set of ordinary differential equations. We
consider an isotropic initial condition so that
$P_+(i_1,0)=P_-(i_1,0)=\delta_{i_1,0}/(2d)$. This directly leads to
$G_+(z,0)=G_-(z,0)=1/(2d)$, $G_0(z,0)=1-1/d$ and to the vanishing of
their derivatives: $G_+'(z,0)=G_-'(z,0)=G_0'(z,0)=0$.  Note that
normalization of probability imposes $G(1,t)=1$. Furthermore,
Eqs. \eqref{eq:gflatticegp}-\eqref{eq:gflatticeg0} at $z=1$ leads to
\begin{equation}\label{eq:ICG}
    G_+(1,t)=G_-(1,t)=\frac 1 {2d};\quad G_0(1,t)=1-\frac1 d\;.
\end{equation}

To calculate $G''(1,t)$, we take twice the derivative with respect to $z$ of both sides of Eqs. \eqref{eq:gflatticegp}-\eqref{eq:gflatticeg0} and set $z=1$. The first derivative yields
\begin{align}
    \frac{\partial}{\partial t}G'_+(1,t)=&pG_+(1,t)-\alpha G_+'(1,t)+\frac{\alpha}{2d}G'(1,t)\;, \label{eq:G'+}\\
    \frac{\partial}{\partial t}G'_-(1,t)=&-pG_-(1,t)-\alpha G_-'(1,t)+\frac{\alpha}{2d}G'(1,t)\;, \label{eq:G'-}\\
    \frac{\partial}{\partial t}G'_0(1,t)=&-\alpha G_0'(1,t)+\alpha\frac{2d-2}{2d}G'(1,t)\;.\label{eq:G'0}
\end{align}
By symmetry, $G'(1,t)=\langle i_1(t)\rangle=0$. Using~Eq.\eqref{eq:ICG}, we can solve Eqs~\eqref{eq:G'+}-~\eqref{eq:G'0} to get 
\begin{equation}
    G'_+(1,t)=-G'_-(1,t)=\frac{p}{2d\alpha} \left[1-e^{-\alpha t}\right],\qquad G'_0(1,t)=0\;. 
\end{equation}

The second derivative with respect to $z$ then gives, for $z=1$,
\begin{equation}
    \frac{\partial}{\partial t}G''(1,t)=2p[G_+'(1,t)-G_-'(1,t)+G_-(1,t)]=\frac{2p^2}{d\alpha}[1-\exp(-\alpha t)]+\frac{p}{d}\;. \label{eq:latticegfddg}
\end{equation}
Using the initial condition $G''(1,0)=0$, the integration of Eq. \eqref{eq:latticegfddg} leads to
\begin{equation}
    \langle\Delta i_1^2(t)\rangle=G''(1,t)+G'(1,t)=\frac{2p^2}{d\alpha}\left(1+\frac{\alpha}{2p}\right)t+\frac{2p^2}{d\alpha^2}[\exp(-\alpha t)-1]\;. \label{eq:MSDlatticeRTP}
\end{equation}
Equation~\eqref{eq:MSDlatticeRTP} predicts the MSD at all times of an RTP on a $d$-dimensional lattice. Note that, as $t\to 0$, the motion of the particle is diffusive, with a diffusion constant $p/(2d)$. This diffusive regime stems from the stochasticity of the hops along the lattice. As $t$ increases, the dynamics transitions into a ballistic regime, typical of an active particle. At large times, when $t\gg \alpha^{-1}$, the tumbles make the large-scale dynamics diffusive. We then recover Eq.~\eqref{eq:MSDlatticeRTPlarget}, using $\langle\Delta i^2(t)\rangle=d\langle\Delta i_1^2(t)\rangle$.
\end{document}